\documentclass[11pt,a4paper]{article}\pdfoutput=1 

\usepackage{jheppub}
\usepackage{hyperref}
\usepackage{booktabs}
\usepackage{xspace}
\usepackage[T1]{fontenc}
\usepackage{xcolor}
\usepackage[toc,page]{appendix}

\newcommand{\noun}[1]{{\tt #1}}
\newcommand{\sss}{\scriptscriptstyle }
\newcommand{\POWHEG}{\noun{POWHEG}}
\newcommand{\POWHEGBOX}{\noun{POWHEG BOX}}

\newcommand{\MINLO}{\noun{MiNLO}}

\newcommand{\HWJMINLO}{\noun{HWJ-MiNLO}}
\newcommand{\HWJMINLOLHE}{\noun{HWJ-MiNLO(LHE)}}
\newcommand{\HWJMINLOPS}{\noun{HWJ-MiNLO(Pythia8)}}

\newcommand{\HNNLOPS}{\noun{HNNLOPS}} 
\newcommand{\HVNNLO}{\noun{HVNNLO}} 
\newcommand{\HVNNLOPS}{\noun{HVNNLOPS}} 
\newcommand{\NNLO}{\textrm{NNLO}}
\newcommand{\HWNNLOLHE}{\noun{HW-NNLOPS(LHE)}} 
\newcommand{\HWNNLOPS}{\noun{HW-NNLOPS}}
\newcommand{\HWNNLOPSpyth}{\noun{HW-NNLOPS(Pythia8)}} 

\newcommand{\DYNNLOPS}{\noun{DYNNLOPS}}
 
\newcommand{\NNLOPS}{\noun{NNLOPS}}
\newcommand{\NLOPS}{\noun{NLOPS}}
\newcommand{\PYTHIA}[1]{\noun{Pythia{#1}}}
\newcommand{\MCatNLO}{\noun{MC@NLO}}

\newcommand{\HW}{\textrm{HW}}
\newcommand{\HZ}{\textrm{HZ}}
\newcommand{\HV}{\textrm{HV}}

\newcommand{\FASTJET}{\noun{FastJet}}

\newcommand{\Kr}{K_{\scriptscriptstyle \mathrm{R}}}
\newcommand{\Kf}{K_{\scriptscriptstyle \mathrm{F}}}

\newcommand{\pt}{p_{\scriptscriptstyle \mathrm{t}}}

\newcommand{\ptw}{p_{\scriptscriptstyle \mathrm{t,W}}}
\newcommand{\ptwh}{p_{\scriptscriptstyle \mathrm{t,HW}}}
\newcommand{\ptjone}{p_{\scriptscriptstyle \mathrm{t,j_{1}}}}
\newcommand{\ptjtwo}{p_{\scriptscriptstyle \mathrm{t,j_{2}}}}

\definecolor{darkgreen}{rgb}{0,0.8,0}
\definecolor{darkpurple}{rgb}{0,0.5,0.5}
\definecolor{darkblue}{rgb}{0,0,0.7}
\definecolor{darkred}{rgb}{0.5,0,0.0}
\definecolor{darkorange}{rgb}{0.8,0.4,0.0}
\definecolor{green}{rgb}{0.0,0.8,0.4}

\newcommand{\thetacs}{\theta^*}
\newcommand{\phics}{\phi^*}
\newcommand{\born}{\Phi_{\sss B}}
\newcommand{\PhiHW}{\Phi_{\sss \HW^*}} 
\newcommand{\PhiHWsimp}{\Phi_{\sss \HW}}
\newcommand{\yhw}{y_{\sss \HW}}
\newcommand{\dyhw}{\Delta y_{\sss \HW}}
\newcommand{\pth}{p_{t,\sss \textrm{H}}}

\newcommand{\mll}{m_{\sss \ell\nu}}
\newcommand{\all}{a_{\sss \ell\nu}}

\usepackage[mathscr]{euscript}

\preprint{\\\\CERN-PH-TH/2016-049 \\ LAPTH-010/16 \\ OUTP-16-04P}

\title{{NNLOPS accurate associated HW production}}

\author[a]{William Astill,} \author[a]{Wojciech Bizo\'n,}
\author[a,b]{Emanuele Re,} \author[c]{Giulia Zanderighi\footnote{On
    leave from Rudolf Peierls Centre for Theoretical Physics,
    University of Oxford, 1 Keble Road, UK}}

\affiliation[a]{Rudolf Peierls Centre for Theoretical Physics,
University of Oxford, 1 Keble Road, UK} 

\affiliation[b]{LAPTh, Universit\'e Savoie Mont Blanc, CNRS, B.P.110,
  Annecy-le-Vieux F-74941, France}

\affiliation[c]{Theoretical Physics Department, CERN, Geneva, Switzerland}

\emailAdd{william.astill@worc.ox.ac.uk}
\emailAdd{wojciech.bizon@physics.ox.ac.uk}
\emailAdd{emanuele.re@lapth.cnrs.fr}
\emailAdd{giulia.zanderighi@cern.ch}

\abstract{We present a next-to-next-to-leading order accurate
  description of associated \HW{} production consistently matched to a
  parton shower. The method is based on reweighting events obtained
  with the \HW{} plus one jet NLO accurate calculation implemented in
  \POWHEG{}, extended with the \MINLO{} procedure, to reproduce NNLO
  accurate Born distributions. Since the Born kinematics is more
  complex than the cases treated before, we use a parametrization of
  the Collins-Soper angles to reduce the number of variables required
  for the reweighting.  We present phenomenological results at 13 TeV,
  with cuts suggested by the Higgs Cross Section Working Group.}


\begin{document}
\maketitle \flushbottom

\section{Introduction}
\label{sec:intro}

After the discovery of the Higgs boson in Run
I~\cite{Aad:2012tfa,Chatrchyan:2012xdj}, one of the main tasks of the
ongoing LHC Run II is to perform accurate measurements of Higgs
properties. This will be done by a thorough investigation of all Higgs
production and decay modes. Higgs boson production in association with
a boson (\HV{}) is the third largest Higgs production mode and so far
has been studied in Run I in different channels, including $b\bar
b$~\cite{Aad:2014xzb,Chatrchyan:2013zna},
$WW^*$~\cite{TheATLAScollaboration:2013hia,CMS:zwa}, and $\tau
\tau$~\cite{CMS:ckv}. Furthermore, for Higgs production in association
with a $Z$ boson, it has been used to set bounds on invisible Higgs
decay modes~\cite{Chatrchyan:2014tja}. Because of the largest
branching ratio of Higgs to bottom quarks, so far the best
significance was found in this channel, by both ATLAS (1.4\,$\sigma$
significance) and CMS (2.2\,$\sigma$ significance).
It is expected that these results will quickly improve in Run II, both
because of the increased luminosity and the higher energy.  Higgs to
bottom quarks is notably difficult because of the very large QCD
background from $g\to bb$, hence it was suggested that associated
production is best studied in a boosted
regime~\cite{Butterworth:2008iy}.  When boosted cuts are applied this
channel becomes one of the most promising places to constrain the
bottom Yukawa coupling.

In ref.~\cite{Brein:2003wg} the inclusive \HV{} ($V=W, Z$) cross
section was computed at NNLO.
In refs.~\cite{Ferrera:2011bk,Ferrera:2014lca} a fully differential
NNLO calculation of \HV{} including all Drell-Yan type contributions
has been presented.
The impact of top-quark loops at this perturbative order has also been
investigated in ref.~\cite{Brein:2011vx}.  In
ref.~\cite{Ferrera:2013yga} NLO corrections to the $H \to bb$ decay
were combined with the NNLO corrections to the production. NLO
electroweak corrections are also known~\cite{Ciccolini:2003jy,Denner:2011id} and
available in the public code \textsc{HAWK}~\cite{Denner:2014cla}. Recently, in
ref.~\cite{Campbell:2016jau} a NNLO calculation of \HV{} was
presented, that includes both Drell-Yan type and top Yukawa
contributions, and that includes decays of the vector bosons and of
the Higgs boson to $b\bar b, \gamma \gamma, W W^*$.

In ref.~\cite{Ferrera:2011bk} it was shown that, while NNLO
corrections to the inclusive \HW{} cross section are tiny, of the
order of 1-2\%, the impact of NNLO corrections can increase
substantially at the LHC when cuts are imposed on the decay products
or when jet-veto criteria are applied.
Since a jet-veto can have a large impact on the size of higher-order
corrections, it should be modelled as accurately as possible. In an
NNLO calculation, however, a jet is made up of only one or two
partons, and no large all-order logarithms are accounted for. Although
in this particular case large logarithms can be resummed quite
precisely (for instance using the approaches of
refs.~\cite{Banfi:2012jm} or~\cite{Becher:2014aya}), it is often very
useful, and at times needed, to model such effects by means of a
fully-differential simulation, where large logarithms are resummed
(although with limited logarithmic accuracy) by a parton shower
algorithm. The precision required for LHC studies also demands that at
least the NLO corrections be included in such event generation tools,
providing therefore predictions where NLO effects are matched to
parton showers (\NLOPS). Thanks to the various implementations of the
\MCatNLO{}~\cite{Frixione:2002ik} and \POWHEG{}~\cite{Nason:2004rx}
algorithms such tools are now routinely used by experimentalists and
theorists.

More specifically, the QCD NLO calculation of associated Higgs
production (\HV{}) was matched to parton showers with the \MCatNLO{}
method~\cite{Frixione:2005gz}, and, more recently, also using
\POWHEG{}~\cite{Luisoni:2013kna}. Ref.~\cite{Luisoni:2013kna} also
contains \NLOPS{} results for $\HV{}+1$ jet, and a merging of the
\HV{} and \HV{}+jet \NLOPS{} simulations, obtained with the so-called
``Multiscale improved NLO'' approach (\MINLO{} in the
following).\footnote{A merging of \HZ{} and \HZ{} + one jet was also
  achieved recently using a merging scale to separate the zero and
  one-jet regions~\cite{Goncalves:2015mfa}.} The \MINLO{} approach was
formulated in ref.~\cite{Hamilton:2012np} and subsequently refined in
ref.~\cite{Hamilton:2012rf}. In the latter work it was shown that for
processes where a colorless system $X$ is produced in a hadronic
collision, one can simulate with \NLOPS{} accuracy both $X$ and $X+1$
jet production simultaneously, without introducing any external
merging scale. In refs.~\cite{Hamilton:2012rf,Hamilton:2013fea} it was
then shown that with a merged generator of $X$ and $X+1$ jet, and the
NNLO computation for $X$ production, one can build an NNLO+parton
shower accurate generator (\NNLOPS{} from now on) for $X$
production. This approach was used to build \NNLOPS{} accurate
generators for Higgs via gluon fusion~\cite{Hamilton:2013fea} and
Drell Yan production~\cite{Karlberg:2014qua}.
Recently, the \MINLO{} method was extended
further~\cite{Frederix:2015fyz} so that the one can merge even three
units of multiplicity while preserving NLO accuracy.
The construction of these \NNLOPS{} generators based on \MINLO{}
relies on a reweighting which is differential in the variables
describing the inclusive $X$-production Born phase space. For Higgs
production this amounts to a one-dimensional reweighting in the Higgs
rapidity, while for Drell Yan production a three-dimensional
reweighting has been used.

In this paper, we use the aforementioned \MINLO{}-based approach to
match the results obtained in ref.~\cite{Luisoni:2013kna} for $\HW{} +
1$ jet production, to the exact NNLO QCD computation of \HW{}
presented in ref.~\cite{Ferrera:2011bk}, thereby obtaining the first
\NNLOPS{} accurate results for \HW{} production, including leptonic
decays of the $W$ boson. We remind the reader that, as in
ref.~\cite{Ferrera:2011bk}, we only include contributions where the
Higgs boson is radiated off a vector boson: top Yukawa contributions,
i.e. contributions from diagrams containing a top-quark loop radiating
an Higgs boson, have not been included in this work.
Since the Born phase-space for $H\ell\nu$ production involves six
variables, one would need to carry out a six-dimensional reweighting,
which is currently numerically unfeasible.  We will describe in the
core of the paper how we deal with this problem.

The paper is organized as follows. In Sec.~\ref{sec:method} we outline
our method, and discuss in particular the treatment of the
multi-dimensional Born phase space. In Sec.~\ref{sec:practical} we
give all details about our practical implementation. In
Sec.~\ref{sec:validation} we validate our results, while in
Sec.~\ref{sec:pheno} we present phenomenological results with cuts
suggested for the writeup of the fourth Higgs Cross Section working
group report. We conclude in Sec.~\ref{sec:conclu}.  In
App.~\ref{App:uncertainties} we give few more details about the scale
variation uncertainties of the results.

\section{Outline of the method}
\label{sec:method}

The method we use in this work is based on achieving \NNLOPS{}
accuracy by reweighting Les Houches events produced by the
\MINLO{}-improved \POWHEG{} \HW{} plus one jet generator
(\HWJMINLO{}). Each event, with a given weight, contains a final state
made of the colorless system (the Higgs boson and the lepton pair from
the $W$ boson) and 1 or 2 additional light QCD partons. \NNLOPS{}
accuracy is obtained by an appropriate rescaling of the original
weight associated to each event. As described in detail in
refs.~\cite{Hamilton:2013fea,Karlberg:2014qua}, the rescaling must be
differential in the variables describing the Born kinematics of the
colorless system.  Concretely, for each event one computes the Born
variables using the kinematics of the colourless partons in the event
kinematics, as is.  Using these observables, a rescaling factor for
each weight is computed. In its simplest form, the rescaling factor
can be written as
\begin{equation}
  \label{eq:W}
  {\cal W}(\born{}) = \frac{\frac{d\sigma^{\textrm{NNLO}}}{d\born{}}}{\frac{d\sigma^{\MINLO}}{d\born{}}}\,, 
\end{equation}
where $\frac{d\sigma^{\textrm{NNLO}}}{d\born{}}$
($\frac{d\sigma^{\MINLO}}{d\born{}}$) is a multi-differential
distribution obtained at pure NNLO level (using \HWJMINLO{} events),
and $\born{}$ denotes the Born phase space.

It is clear that, by construction, Born variables will be described
with NNLO accuracy. Furthermore, since the \HWJMINLO{} is NLO accurate
for distributions inclusive on all radiation, it is straightforward to
prove (along the lines of the proofs presented in
refs.~\cite{Hamilton:2013fea,Karlberg:2014qua}) that this rescaling
does not spoil the NLO accuracy of \HWJMINLO{} generator.  As a
consequence of these two facts, after rescaling, one obtains full NNLO
accuracy for \HW{}.

One might worry that once events undergo a parton shower, the NNLO
accuracy might be lost. It is however easy to see that this is not the
case: the second emission is generated by \POWHEG{} precisely in such
a way as to preserve the NLO accuracy of 1-jet observables. Hence the
first emission generated by the parton shower is the third one,
i.e. the effect of the parton shower starts at ${\cal O}(\alpha_s^3)$,
and is therefore beyond NNLO.

In the present case, the Born kinematics is fully specified by six
independent variables. For instance one can choose the rapidity of the
$\HW{}$-system ($\yhw{}$), the difference in rapidity between the Higgs
and the $W$ boson ($\Delta \yhw{}$), the Higgs transverse momentum
($\pth{}$), the dilepton pair invariant mass ($\mll{}$) and two
angular variables. A convenient standard choice for the angular
variables is to use the Collins-Soper angles~\cite{Collins:1977iv}
defined as follows. One considers a boost from the laboratory frame to
the rest frame of the $W$ boson (the ${\cal O}'$ frame). Using the
positive and negative rapidity beam momenta, respectively $p'_A$ and
$p'_B$ in ${\cal O}'$, one defines a $z$-axis in this frame such that
it bisects the angle between $p'_A$ and $-p'_B$. One then introduces a
transverse unit vector $\hat q_T$, orthogonal to the $z$ axis and
lying in the ($p'_A, p'_B$) plane, pointing away from $p'_A+p'_B$. The
Collins-Soper angles are defined as the polar angle $\thetacs$ of the
lepton momentum $l'$ in ${\cal O}'$ with respect to the $z$-axis
($\vec l' \cdot \hat z = |l'|  \cos \thetacs$) and the azimuthal angle
$\phics$ of $l'$ ($\vec l' \cdot \hat q_T = |l'| \sin \thetacs\cos
\phics$).

Since the decay of a massive spin one particle is at most quadratic in
the lepton momentum $\vec l'$ in the frame ${\cal O}'$, one can
parametrize the angular dependence in terms of the nine spherical
harmonic functions $Y_{lm}(\thetacs,\phics)$ with $l\le 2$ and $|m|\le
l$.
This can be understood from the observation that the decay of a
massive spin one particle is associated to 9 degrees of freedom (the
spin-density matrix is a 3x3 matrix).  One of these coefficients is
then fixed by the normalisation of the cross section, so that eight
independent coefficients are sufficient to parametrize the angular
dependence.  As it is done in the case of Drell-Yan, it is convenient
to introduce the following parametrisation for the angular dependence,
\begin{eqnarray}
\label{eq:sigma}
\frac{d\sigma}{d\born{}}  &=& 
\frac{d^6\sigma}{d\yhw{}\, d\Delta \yhw{}\, d\pth{}\, d\mll{}\, d\cos \thetacs d\phics} \nonumber \\
&=&  \frac{3}{16\pi}  \left ( 
\frac{d \sigma}{d\PhiHW}(1+\cos^2\thetacs) + \sum_{i=0}^{7} A_i(\PhiHW ) f_i(\thetacs, \phics)
\right)\,,  
\end{eqnarray}
where we introduced for simplicity the four dimensional phase space of the
$\HW{}^*$ system, $\PhiHW = \{ \yhw{}, \Delta \yhw{}, \pth{}, \mll{}
\}$
and  $\frac{d \sigma}{d\PhiHW}$ corresponds to the fully differential
cross section integrated just over the Collins-Soper angles. The functions $f_i(\thetacs, \phics)$ are essentially given
by spherical harmonics
\begin{equation}
\label{eq:f}
\begin{aligned}
f_0(\thetacs,\phics) &= \left(1-3\cos^2\thetacs\right)/2\,,\qquad \\
f_2(\thetacs,\phics) &= (\sin^2\thetacs \cos2\phics)/2\,, \\
f_4(\thetacs,\phics) &= \cos\thetacs\,, \\ 
f_6(\thetacs,\phics) &= \sin 2\thetacs \sin \phics\,, \\
\end{aligned}
\begin{aligned}
f_1(\thetacs,\phics) &= \sin2\thetacs \cos\phics\,, \\
f_3(\thetacs,\phics) &= \sin\thetacs \cos\phics\,, \\ 
f_5(\thetacs,\phics) &= \sin\thetacs \sin \phics\,, \\ 
f_7(\thetacs,\phics) &= \sin^2\thetacs \sin 2\phics\,.  
\end{aligned}
\end{equation}
They have the property that their integral over the solid angle
$d\Omega = d\cos\thetacs d\phics$ vanishes.

Since the angular dependence is fully expressed in terms of the
$f_i(\thetacs,\phics)$ functions, the coefficients of the expansion
$A_i(\PhiHW)$ are functions only of the remaining kinematical variables
$\PhiHW$.
The coefficients $A_i(\PhiHW )$ can then be extracted using
orthogonality properties of the spherical harmonics.  We find
\begin{equation}
\begin{aligned}
A_0( \PhiHW ) &= 4\, ({d \sigma}/{d\PhiHW}) -  \langle 10 \cos^2 \thetacs \rangle\,, \quad  \\
A_2( \PhiHW ) &= \langle 10 \sin^2\thetacs \cos 2\phics \rangle\,, \quad  \\
A_4( \PhiHW  ) &= \langle 4 \cos\thetacs\rangle\,, \quad  \\
A_6( \PhiHW  ) &= \langle 5 \sin2\thetacs \sin\phics\rangle\,, \quad  \\
\end{aligned}
\begin{aligned}
A_1( \PhiHW  )&=  \langle 5 \sin 2\thetacs \cos\phics \rangle\,, \\
A_3( \PhiHW  ) &= \langle 4 \sin\thetacs \cos\phics\rangle\,,\\
A_5( \PhiHW  ) &= \langle 4 \sin\thetacs \sin\phics \rangle\,, \\ 
A_7( \PhiHW ) &= \langle 5 \sin^2\thetacs \sin2\phics\rangle\,,
\end{aligned}
\label{eq:A}
\end{equation}
where the expectation values $\langle f(\thetacs,\phics) \rangle$ are
functions of $\PhiHW$  defined as
\begin{equation}
\label{eq:average}
\langle f(\thetacs,\phics) \rangle = 
\int {d\cos\thetacs d\phics} \frac{d\sigma}{d\born{}}f(\thetacs,\phics)\,. 
\end{equation}

Hence, in order to compute both the numerator and denominator in
eq.~\eqref{eq:W}, as required for the reweighting, we can use
eq.~\eqref{eq:sigma} with the angular functions defined in
eq.~\eqref{eq:f} and the coefficients computed using eq.~\eqref{eq:A}.
In summary, by using the Collins-Soper angles one can turn the problem
of computing differential distributions in six variables, into the
determination of nine four-dimensional distributions, i.e.\ $d\sigma
/d\PhiHW$ and the eight distributions $A_i(\PhiHW)$ of
eq.~\eqref{eq:A}.

\section{Practical implementation}
\label{sec:practical}

In the previous section we have outlined the method that we will use
in the following to achieve \NNLOPS{} accuracy. Here, we will provide details about the choices that
we made in our practical implementation, we outline the setup that we
have adopted to present the results of this paper, and we give the
procedure that we used to estimate the theoretical uncertainty.

\subsection{Procedure}

A first consideration is that when using multi-differential
distributions one needs to decide the number of bins in each
distribution. Previous experience suggests that having about 25 bins
per direction is sufficient for practical purposes, hence we will
adopt this choice here. In order to improve the numerical precision,
we find it useful to use bins that contain approximately the same
cross-section, as opposed to bins that are equally
spaced. Practically, we perform (moderate statistics) warm-up runs at
NLO using \HWJMINLO{}. From the differential cross sections obtained
from these runs, we determine the appropriate bins. We then read in
the bin values when performing high-statistic runs to extract the
needed distributions.

We have simplified our procedure by noting that the $\mll{}$ invariant
mass distribution has a flat $K$-factor. This is true even when
examining the $d\sigma/d{\mll{}}$ distribution in different bins of
$\PhiHWsimp = \{ \yhw{}, \Delta \yhw{}, \pth{}\}$.
Therefore, in eq.~\eqref{eq:sigma} we replace $\PhiHW$ with
$\PhiHWsimp$ and in eq.~\eqref{eq:average} we integrate over $m_{\ell
  \nu}$, meaning that instead of having four-dimensional
distributions, we use three-dimensional ones.  This is an
approximation, however we believe that it works extremely well, as
discussed in Sec.~\ref{sec:validation}.

A further point to note is that, as observed already in
ref.~\cite{Hamilton:2013fea}, a reweighting of the form
eq.~\eqref{eq:W} spreads the $\textrm{NNLO/NLO}$ $K$-factor uniformly,
even in regions where the \HW{} system has a large transverse
momentum, i.e. a region that is described equally well by a pure NNLO
\HW{} calculation, or by the \HWJMINLO{} generator. However, it is
also possible to introduce a reweighting that goes smoothly to one in
the regions where both generators have the same accuracy to start
with. In order to do this, one introduces a smooth function of
$p_{\sss T}$, that goes to one at $p_{\sss T}=0$ and that vanishes at
infinity. For instance, one can introduce
\begin{equation}
h(p_{\sss T}) = \frac{(M_{\sss H}+M_{\sss W})^2}{(M_{\sss H}+M_{\sss W})^2+p_{\sss T}^{\; 2}}\,, \label{eq:h_pt}
\end{equation}
to split the cross-section into
\begin{equation}
d\sigma_{\sss A} = d\sigma\, h(p_{\sss T})\,,\qquad  
d\sigma_{\sss B} = d\sigma\, (1-h(p_{\sss T}))\,. 
\end{equation}
One then reweights the \HWJMINLO{} events using 
\begin{eqnarray} 
  \mathcal{W}\left(\PhiHWsimp,\, p_{{\sss
      \mathrm{T}}}\right)&=&h\left(\pt\right)\,\frac{\smallint
    d\sigma^{{\sss
        \mathrm{NNLO\phantom{i}}}}\,\delta\left(\PhiHWsimp-\PhiHWsimp\left(\Phi\right)\right)-\smallint
    d\sigma_{\sss B}^{{\sss
        \mathrm{\MINLO}}}\,\delta\left(\PhiHWsimp-\PhiHWsimp\left(\Phi\right)\right)}{\smallint
    d\sigma_{\sss A}^{{\sss
        \mathrm{\MINLO}}}\,\delta\left(\PhiHWsimp-\PhiHWsimp\left(\Phi\right)\right)}\nonumber\\ &+&\left(1-h\left(\pt\right)\right)\label{eq:NNLOPS-overall-rwgt-factor-1}\,.
\end{eqnarray} 
This reweighting factor preserves the exact value of the NNLO differential cross-section
\begin{eqnarray}
  \left(\frac{d\sigma}{d\PhiHWsimp}\right)^{{\sss
      \mathrm{NNLOPS}}} & = &
  \left(\frac{d\sigma}{d\PhiHWsimp}\right)^{{\sss
      \mathrm{NNLO}}}\,.\label{eq:NNLOPS-NNLOPS-eq-NNLO_0+MINLO_1-1}
\end{eqnarray}

We choose $p_{\sss T}$ to be the transverse momentum of the leading
jet when clustering events with the inclusive $k_T$-algorithm with
$R=0.4$~\cite{Catani:1993hr,Ellis:1993tq}. The reason for this is
choice is that $h(p_T)$ goes to one when no radiation is present,
since the leading jet transverse momentum vanishes. On the contrary,
when hard radiation is present, the transverse momentum of the leading
jet becomes large, $h(p_T)$ goes to zero, and accordingly ${\cal
  W}(\PhiHWsimp,p_{\sss T})$ goes to one.

\subsection{Settings}

We give here a complete description of the setup used for the results
presented in this paper. The specific process studied is
\begin{equation}
  pp\longrightarrow HW^{+} \longrightarrow H\ell^{+}\nu_{\ell}\,, 
\end{equation}
where $\ell^+ = \{e^+, \mu^+\}$.\footnote{When running the code we
  fixed the W boson decay to the electron channel and multiplied the
  result by two to include the muon channel.}  We note that we leave
the Higgs boson in the final state, rather than decaying it.

We used the code \HVNNLO{}~\cite{hvnnlo} to obtain NNLO predictions,
and the \HWJMINLO{} code~\cite{Luisoni:2013kna} implemented in the
\POWHEGBOX{}~\cite{Alioli:2010xd} to produce Les Houches
events.\footnote{As specified in Sec.~\ref{sec:intro}, we have
  neglected contributions where the Higgs boson is produced by a
  top-quark loop.
  This has been achieved by setting the flag \noun{massivetop} to zero
  when running the \HWJMINLO{} program.}
Throughout this work we consider 13 TeV LHC collisions and use the
MMHT2014nnlo68cl parton distribution
functions~\cite{Harland-Lang:2014zoa}, corresponding to a value of
$\alpha_s(M_Z) = 0.118$.  We set $M_W = 80.399 $ GeV and $\Gamma_W =
2.085$ GeV. Furthermore we use $\alpha_{\rm em} = 1/132.3489$ and
$\sin^2\theta_W = 0.2226$. Finally we use $M_H=125$ GeV.
Jets have been constructed using the anti-$k_t$ algorithm with
$R=0.4$~\cite{Cacciari:2008gp} as implemented in
\FASTJET{}~\cite{Cacciari:2005hq,Cacciari:2011ma}.
For \HWJMINLO{} events the scale choice is dictated by the \MINLO{}
procedure; for the NNLO we have used for the central renormalisation
and factorisation scales $\mu_0 = M_H+M_W$.

To shower partonic events we have used
\PYTHIA{8}~\cite{Sjostrand:2007gs} (version 8.185) with the ``Monash
2013''~\cite{Skands:2014pea} tune. 
To define leptons from the boson decays we use the Monte Carlo truth,
i.e. we assume that if other leptons are present, the ones coming from
the $W$ decay can be identified correctly.
To obtain the results shown in the following sections, we have
switched on the ``doublefsr'' option introduced in
ref.~\cite{Nason:2013uba}.
The plots shown throughout the paper have been obtained keeping the
veto scale equal to the default \POWHEG{} prescription.

\subsection{Estimating uncertainties} \label{subsec:Estimating-uncertainties} 

We outline here the procedure that we use to estimate the
uncertainties in our \NNLOPS{} event generator. This procedure is similar
to the one already used in
refs.~\cite{Hamilton:2013fea,Karlberg:2014qua}, but we find it useful
to recall it here for completeness. %
As is standard, the uncertainties in the \HWJMINLO{} generator are
obtained by varying by a factor 2 up and down independently all
renormalisation scales appearing in the \MINLO{} procedure by $\Kr$
(simultaneously) and the factorisation scale by $\Kf$, keeping $1/2
\le \Kr/\Kf \le 2$. This leads to 7 different scale choices given by
\begin{equation}\label{scales}
(\Kr,\Kf)=(0.5,0.5),(1,0.5),(0.5,1),(1,1),(2,1),(1,2),(2,2)\,\text{.}	
\end{equation}
The seven scale variation combinations have been obtained by using the
reweighting feature of the \POWHEGBOX{}. 

For the pure NNLO results the uncertainty band is the envelope of the
same 7-scale variations as used for \HWJMINLO{} uncertainties.
Currently, in the next-to-next-to-leading order computation in
\HVNNLO{}, the only way of doing scale variations is to re-run the
entire program with new scales.  To be more efficient, one can instead
compute the NNLO result at just 3 scale choices for $\mu_F$,
e.g. $(\Kr,\Kf)=(1,0.5),(1,1),(1,2)$, along with pure LO and NLO
results.  One can then use renormalisation group equations to predict
results at different renomalization scales.

For the \NNLOPS{} results, we have first generated a single \HWJMINLO{}
event file with all the weights needed to compute the integrals
$d\sigma_{A/B}^{\scriptscriptstyle \mathrm{MINLO}}/d\born{}$ entering
eq.~(\ref{eq:NNLOPS-overall-rwgt-factor-1}) for all 7 scale choices.

The differential cross-section $d\sigma^{\scriptscriptstyle
  \mathrm{NNLO}}/d\Phi$ was tabulated for each of the seven scale
variation points corresponding to $1/2 \le \Kr'/\Kf' \le 2$.  The
analysis is then performed by processing the \MINLO{} event for given
values of $(\Kr,\Kf)$, and multiplying its weight with the factor
\begin{equation} h\left(\pt\right)\times\,\frac{\smallint
    d\sigma_{(\Kr',\Kf')}^{{\scriptscriptstyle
        \mathrm{NNLO\phantom{i}}}}\,\delta\left(\born{}-\born{}(\Phi)\right)-\smallint
    d\sigma_{B,(\Kr,\Kf)}^{{\scriptscriptstyle
        \MINLO}}\,\delta\left(\born{}-\born{}(\Phi)\right)}{\smallint
    d\sigma_{A,(\Kr,\Kf)}^{{\scriptscriptstyle
        \MINLO}}\,\delta\left(\born{}-\born{}(\Phi)\right)}+\left(1-h\left(\pt\right)\right)\,.\label{eq:master} 
\end{equation}
The central value is obtained by setting $(\Kr,\Kf) = (\Kr',\Kf') =
(1,1)$, while to obtain the uncertainty band we apply this formula for
all the seven $(K_{{\scriptscriptstyle
    \mathrm{R}}},K_{{\scriptscriptstyle \mathrm{F}}})$ and seven
$(K_{{\scriptscriptstyle \mathrm{R}}}^{\prime},K_{{\scriptscriptstyle
    \mathrm{F}}}^{\prime})$ choices. This yields 49 scale variations
in the final \NNLOPS{} accurate events.\footnote{We have checked that
  performing instead a 21-point variation, i.e. doing only a 3-point
  scale variation in the NNLO result, leads in general to only
  moderately smaller uncertainties, as discussed in
  Appendix~\ref{App:uncertainties}.}

As explained in refs.~\cite{Hamilton:2013fea,Karlberg:2014qua}, the
motivation to vary scales in the NNLO and \HWJMINLO{} results
independently is that, in the same spirit of the efficiency
method~\cite{Banfi:2012yh}, we regard uncertainties in the overall
normalisation of distributions as being independent of the respective
uncertainties in the shapes.

\section{Validation}
\label{sec:validation}

\subsection{Validation of the \NNLOPS{} method}
\label{subsec:validation-reweighting}

Our method uses the approximation that the $K$-factor of the dilepton
system invariant mass is flat in the whole phase space.  Hence, we
first discuss how good this approximation is.
 
\begin{figure}[h]
  \centering
  {{ \includegraphics[width=0.48\textwidth]{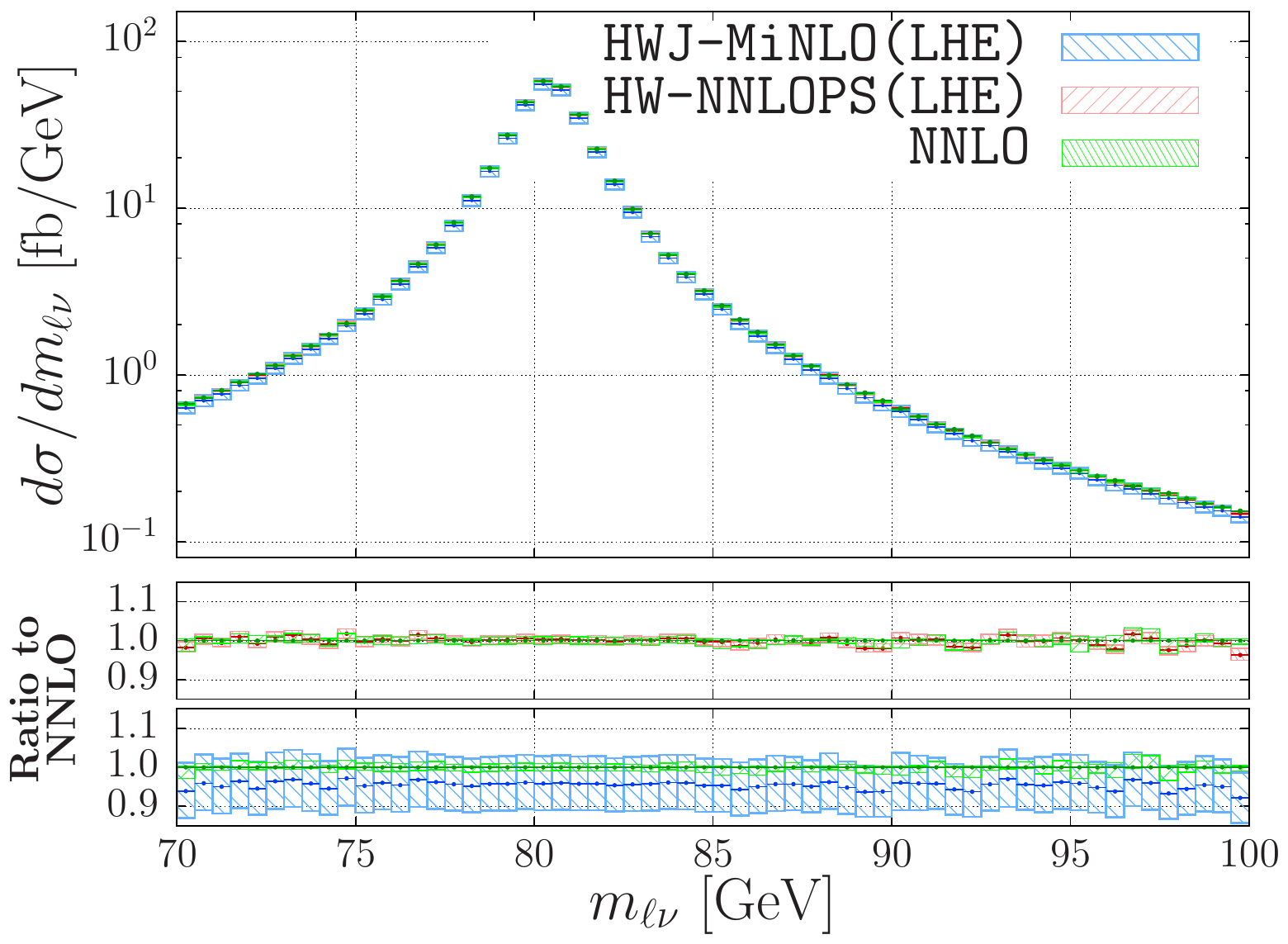} }}
  {{ \includegraphics[width=0.48\textwidth]{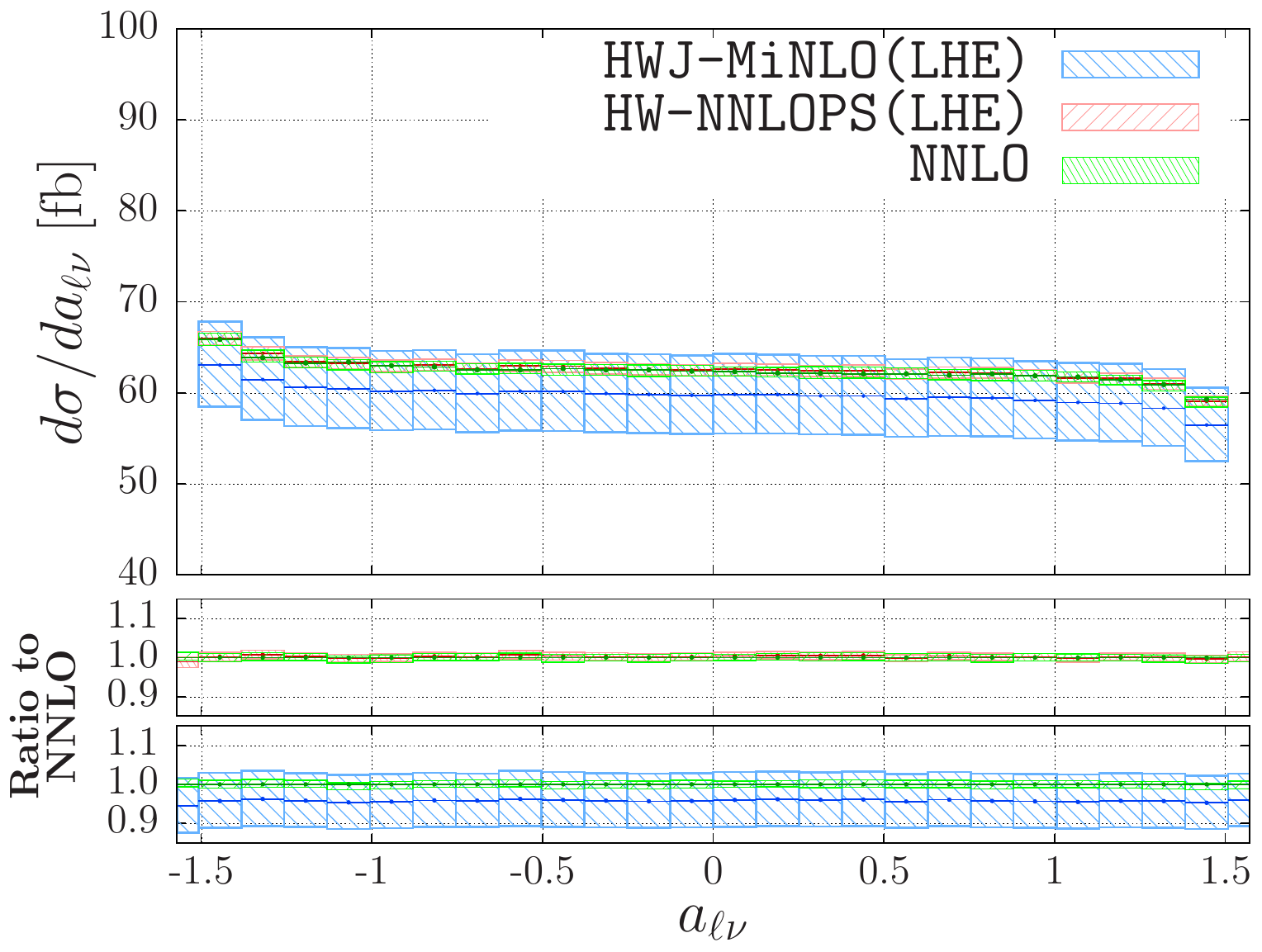} }}
  \caption{Comparison of \HWJMINLOLHE{} (blue), \NNLO{} (green) and \HWNNLOLHE{} (red) for $\mll{}$ (left) and $\all{}$ (right) defined in eq.~\eqref{eq:a_ll}.}
  \label{fig:m_v}
\end{figure}

Figure \ref{fig:m_v} (left) 
shows the distribution of the $(l,\nu)$-invariant mass $m_{\ell \nu}$
integrated over the whole phase space.  The right plot shows the
distribution of
\begin{equation}\label{eq:a_ll}
  \all{} = \arctan\left( \frac{\mll{} - m_{W}}{m_{W}\Gamma_{W}} \right)
\end{equation}
which is constructed in order to flatten the $\mll{}$ distribution.
The upper panels show the predictions from \HWJMINLOLHE{} at pure Les
Houches event (LHE) level, i.e. including NLO and Sudakov effects, but
prior to parton shower (blue), predictions at \HWNNLOLHE{} level,
i.e. including NNLO corrections and Sudakov effects but no parton
shower (red) and \NNLO{} results (green).  The lower panels show the
ratio to the NNLO result. The uncertainty bands are computed as
described in Sec.~\ref{subsec:Estimating-uncertainties}. We notice
that \NNLO{} and \HWNNLOLHE{} predictions agree very well within their
small uncertainty bands. \HWJMINLOLHE{} predictions are about 5\%
lower, but, as expected, the NNLO/NLO $K$-factor is flat over the
whole region. In fact, the distributions have a Breit-Wigner shape,
hence one expects higher-order corrections to affect the shape only
very mildly, if at all.

\begin{figure}[h]
  \centering
  {{ \includegraphics[width=0.48\textwidth]{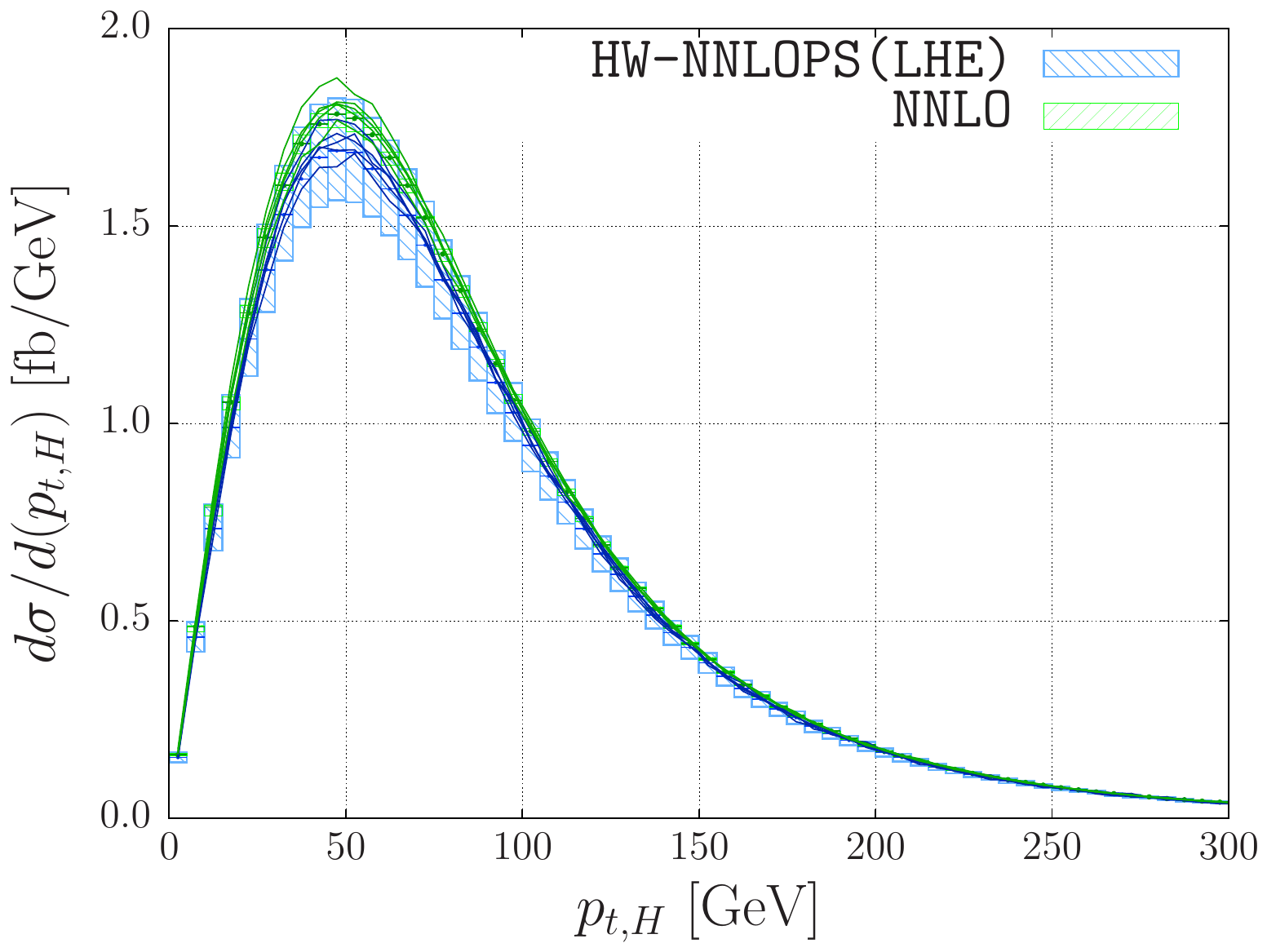} }}
  {{ \includegraphics[width=0.48\textwidth]{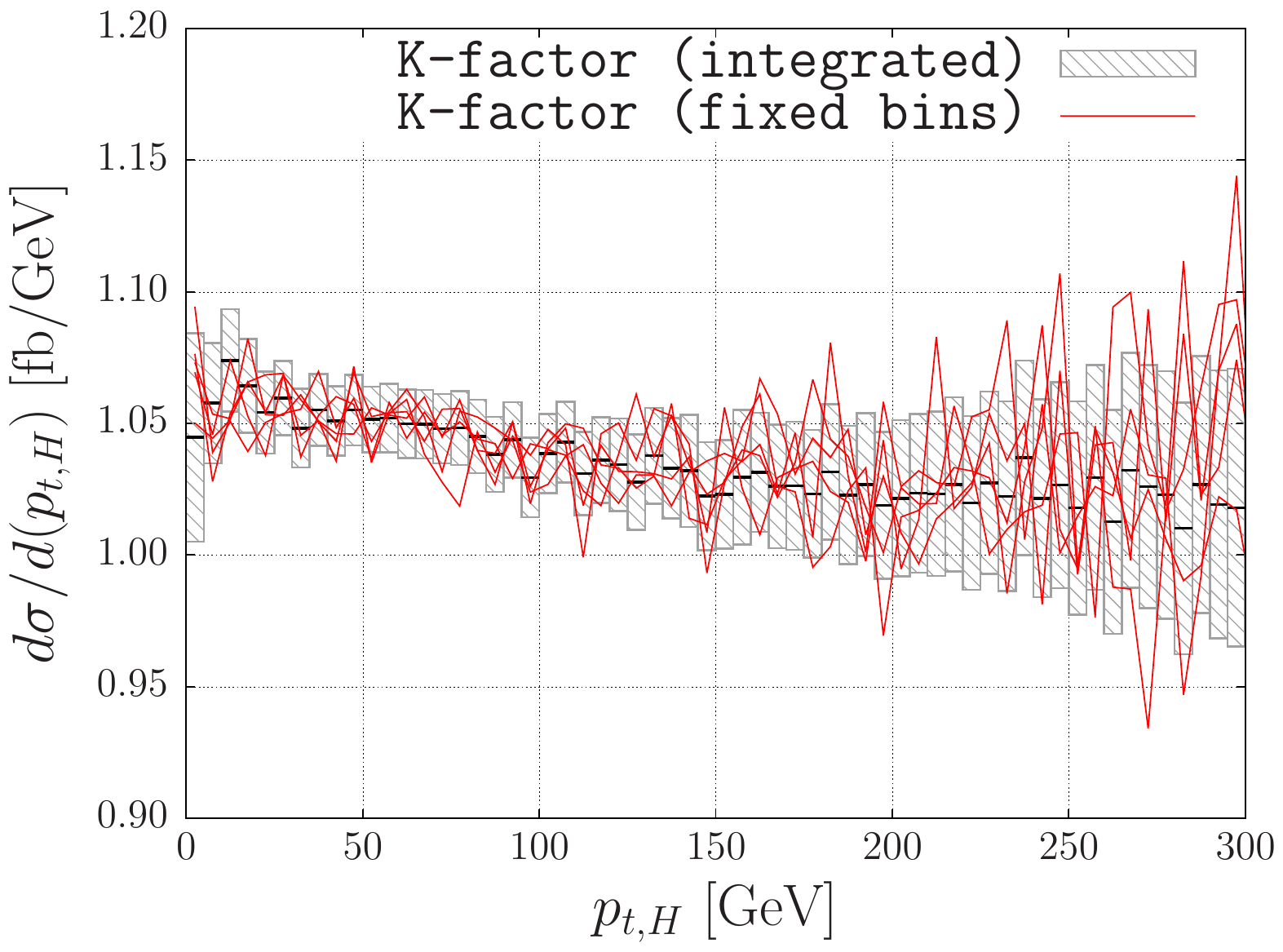} }}
  {{ \includegraphics[width=0.48\textwidth]{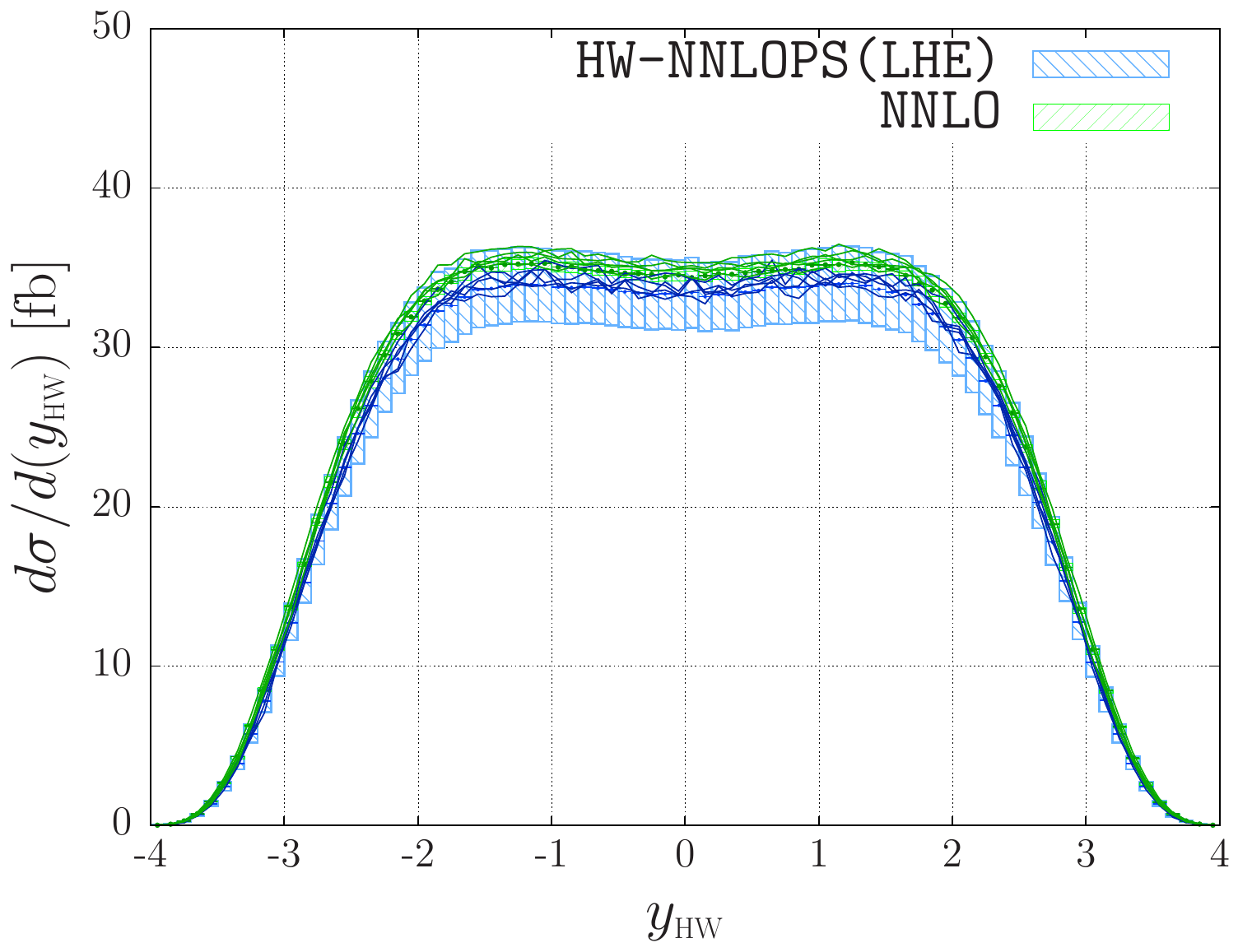} }}
  {{ \includegraphics[width=0.48\textwidth]{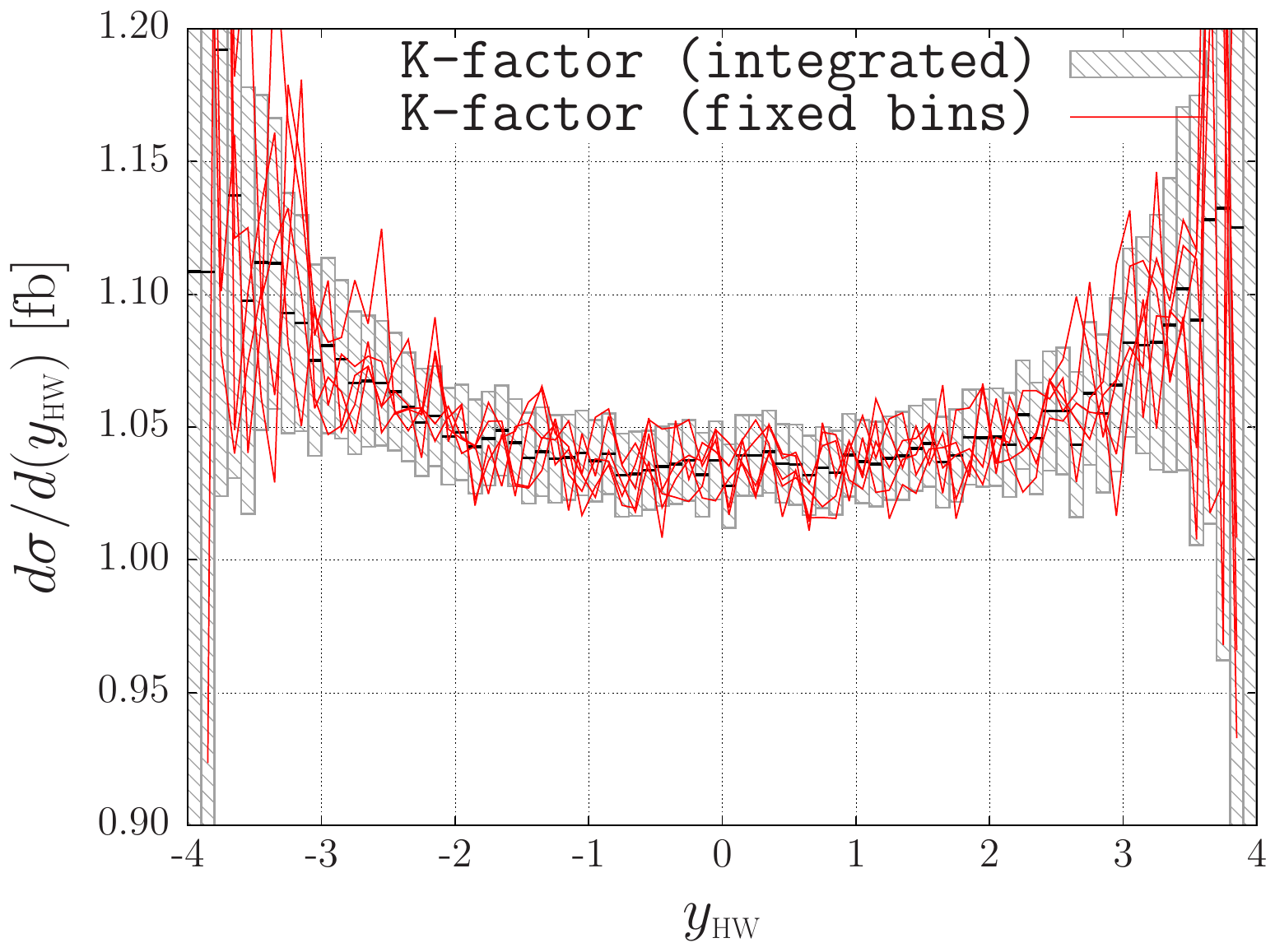} }}
  {{ \includegraphics[width=0.48\textwidth]{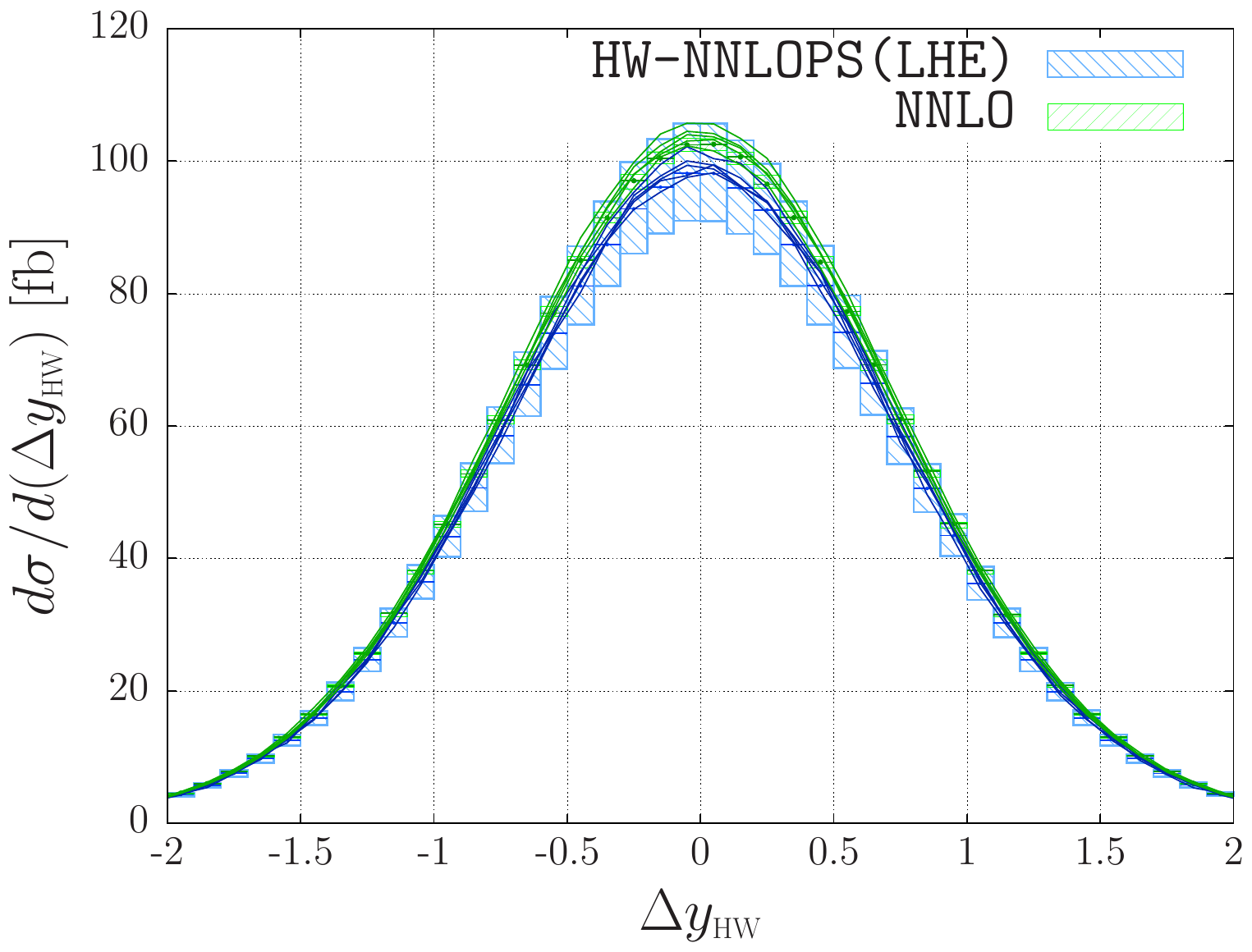} }}
  {{ \includegraphics[width=0.48\textwidth]{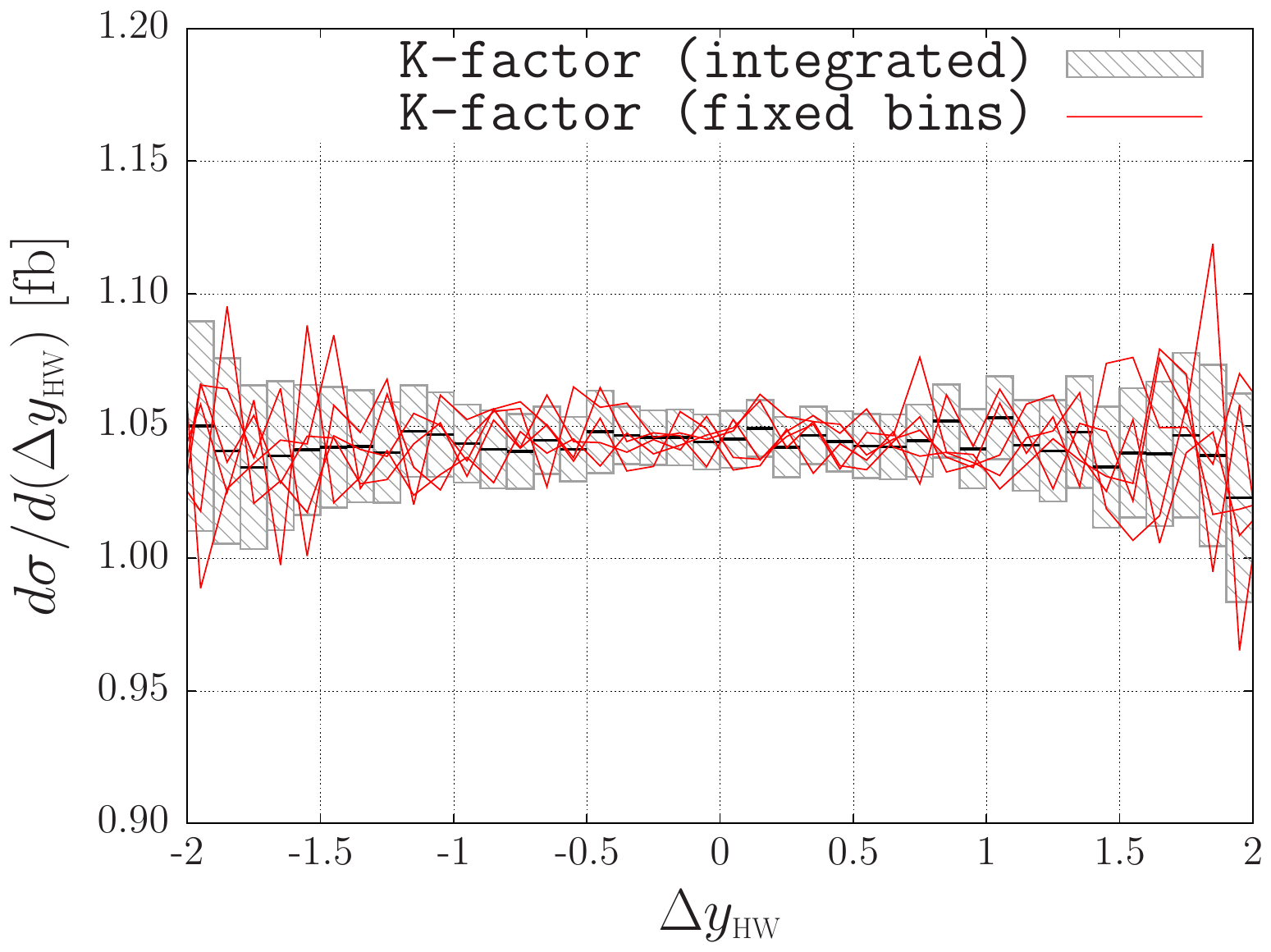} }}
  \caption{Comparison of \HWJMINLOLHE{} (blue) and \NNLO{} (green) for Born variables {$\PhiHWsimp$} chosen to perform
    reweighting.  Left-panel: boxes represent results integrated over
    whole phase space (with theoretical uncertainty), whereas lines come from various $\all{}$ bins
    (as described in the text).  Right-panel: boxes represent the overall
    $K$-factor (integrated over $\all{}$) with statistical
    uncertainty, while lines represent $K$-factors corresponding to various
    $\all{}$ bins (bin 3, 8, 13, 18, 23).}
  \label{fig:val-mv}
\end{figure}

Since our reweighting procedure is differential in all Born variables,
but for $\mll{}$, we need to also verity that the NNLO/NLO $K$-factor
is flat in bins of all other Born variables. This is equivalent to
saying that the NNLO/NLO $K$-factors for all other Born variables
should be the same in every bin in $\mll$ (or equivalently in
$\all{}$). In Fig.~\ref{fig:val-mv} (left)
we show the comparison between \HWJMINLOLHE{} and \NNLO{} for the
three Born variables $\yhw{}$, $\pth{}$ and $\Delta \yhw{}$. Here the
blue and green bands represent the usual theoretical uncertainty.  In
the right panels the black line shows the $K$-factor integrated over
the whole $\all{}$ range and the five red lines show the same
$K$-factor in a fixed $\all{}$ bin.\footnote{For clarity, we show only
  5, rather than all lines. We have verified that the picture does not
  change when all lines are displayed.}  Now, the grey band
corresponds to the statistical uncertainty of the $K$-factor
integrated over the whole $a_{l \nu}$ range, multiplied by a factor 5.
Since we are probing 25 bins in $\all{}$, one expects that the
statistical uncertainty for a particular bin is bigger by
$\sqrt{N_{\textrm{bins}}}$ (we recall that the $\all{}$ distribution
is by construction fairly flat). Therefore this band provides an
estimate of the uncertainty of the $K$-factor on each $\all{}$ bin.
We see that, within statistical
fluctuations, the red lines lie within the grey band. This shows that,
within the statistical uncertainties, the $K$-factor is independent of
the value of $\all{}$.

\begin{figure}[t]
  \centering
  {{ \includegraphics[width=0.48\textwidth]{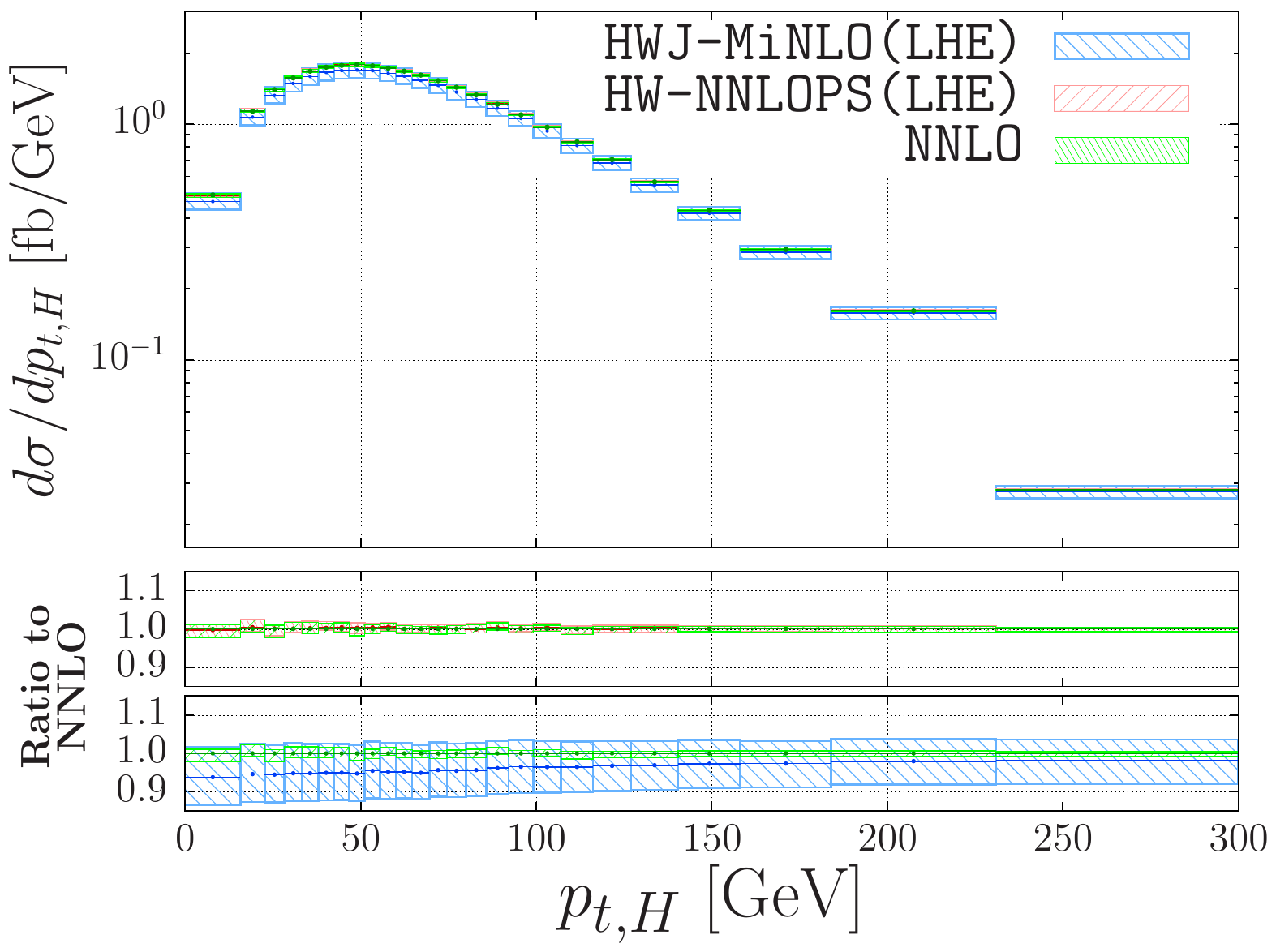} }}
  {{ \includegraphics[width=0.48\textwidth]{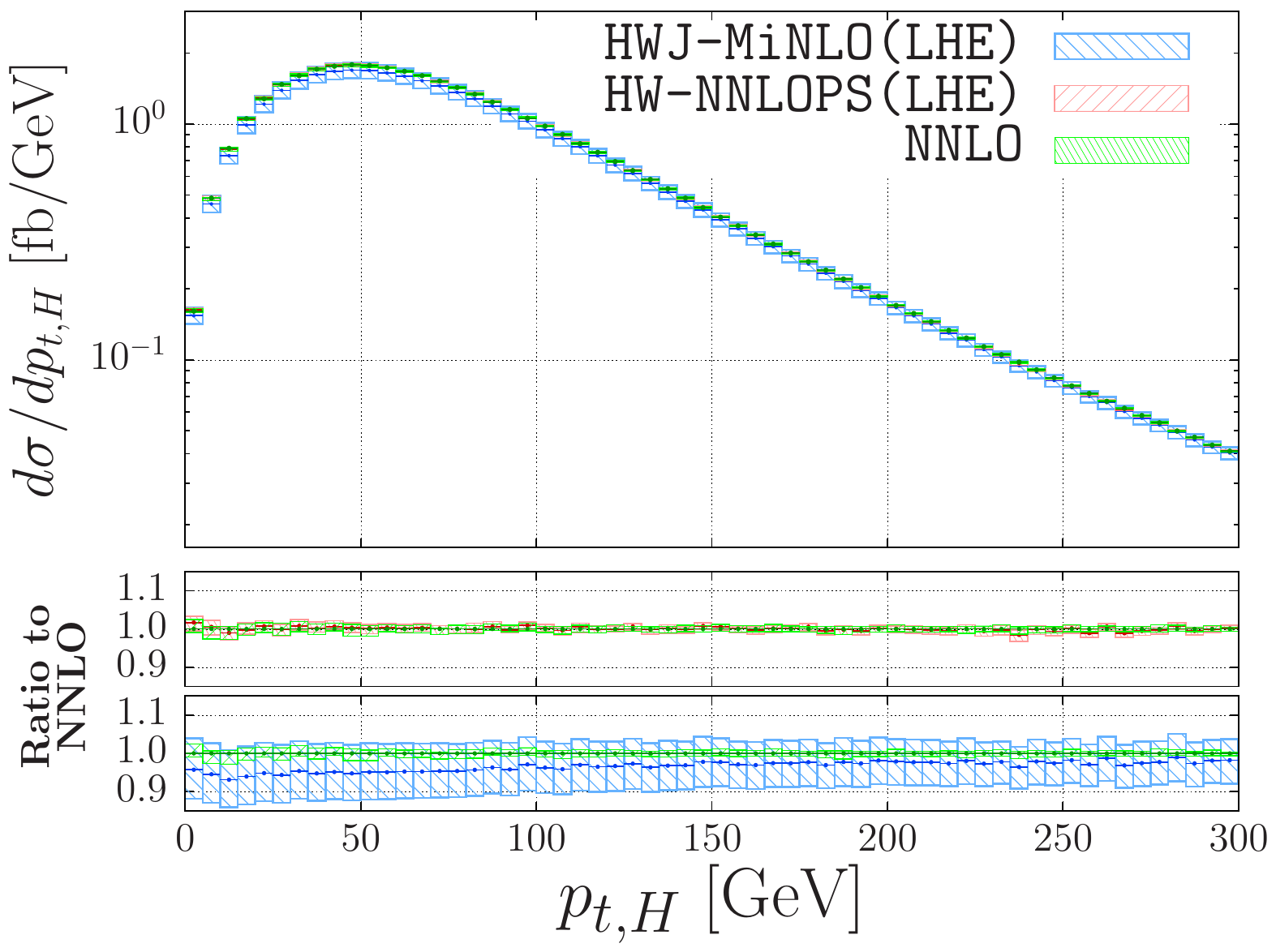} }}
  {{ \includegraphics[width=0.48\textwidth]{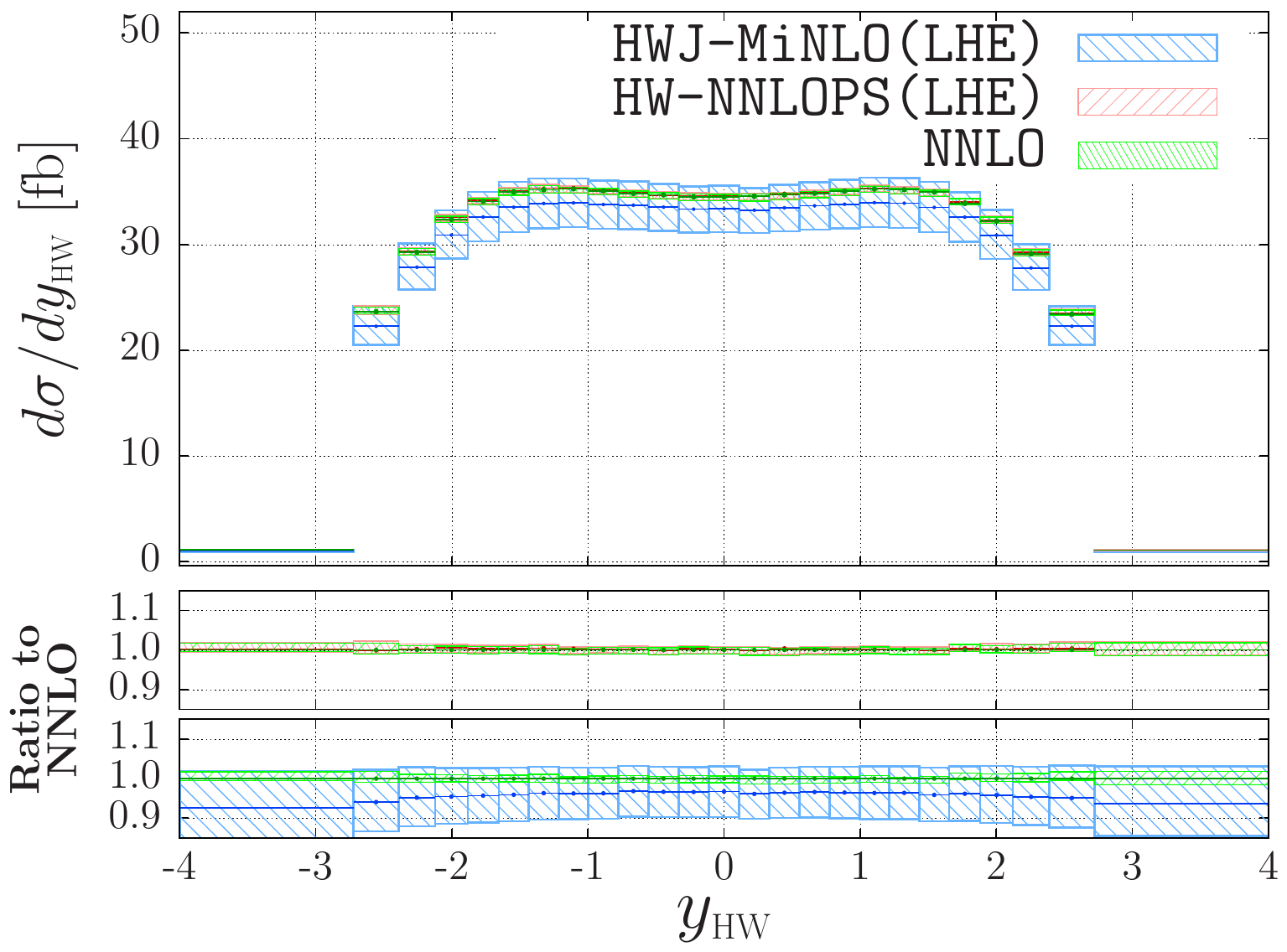} }}
  {{ \includegraphics[width=0.48\textwidth]{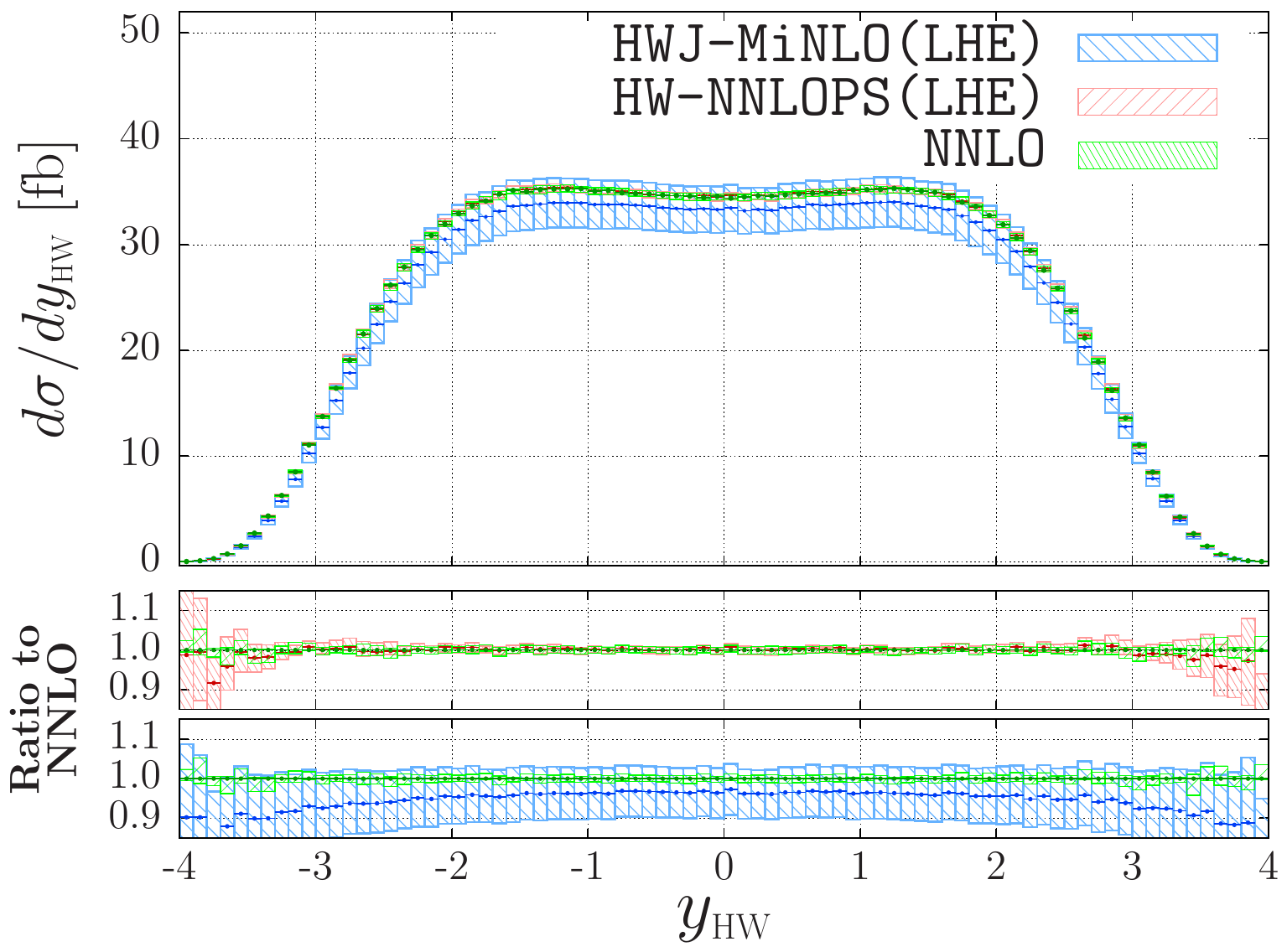} }}
  {{ \includegraphics[width=0.48\textwidth]{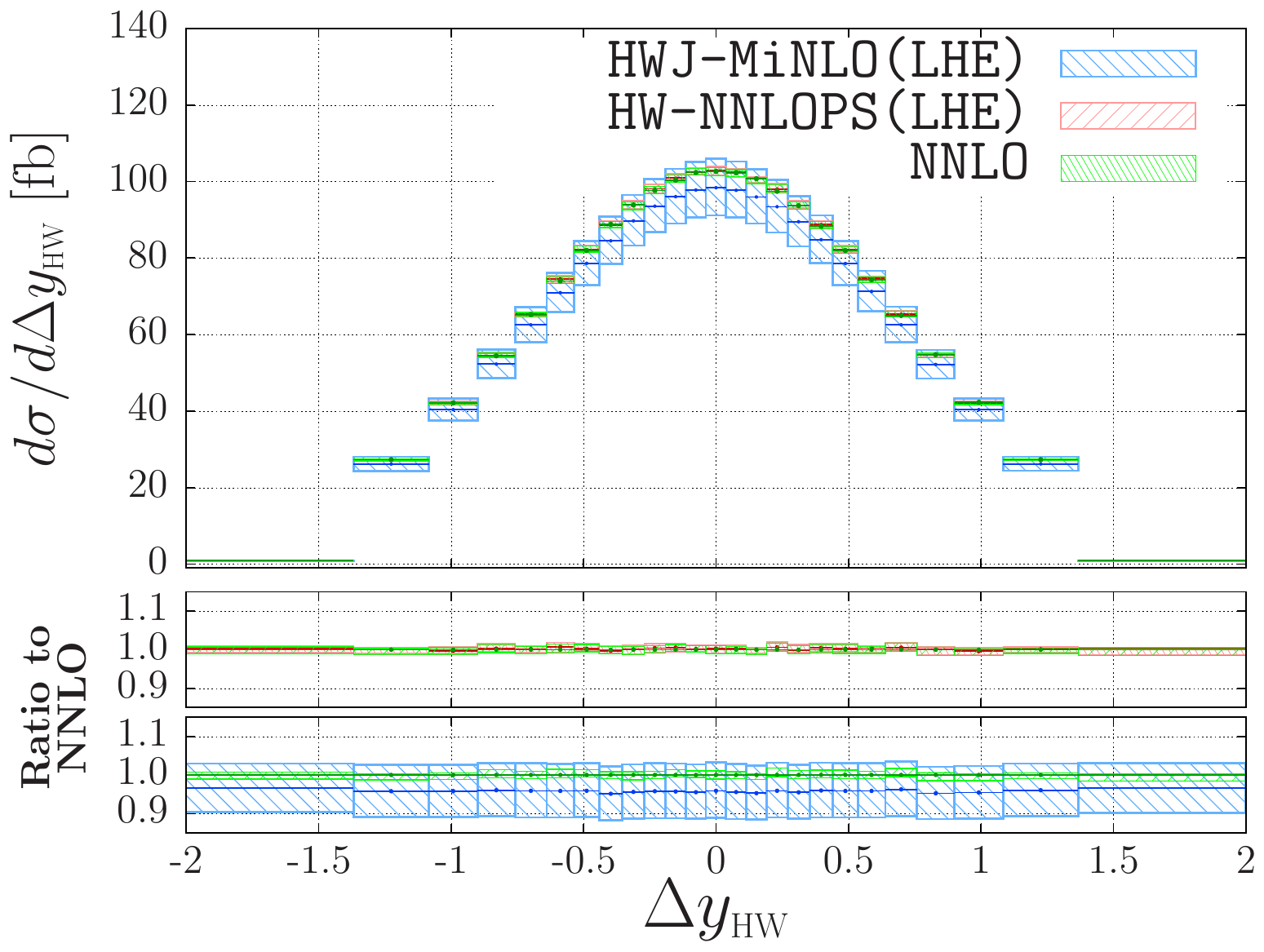} }}
  {{ \includegraphics[width=0.48\textwidth]{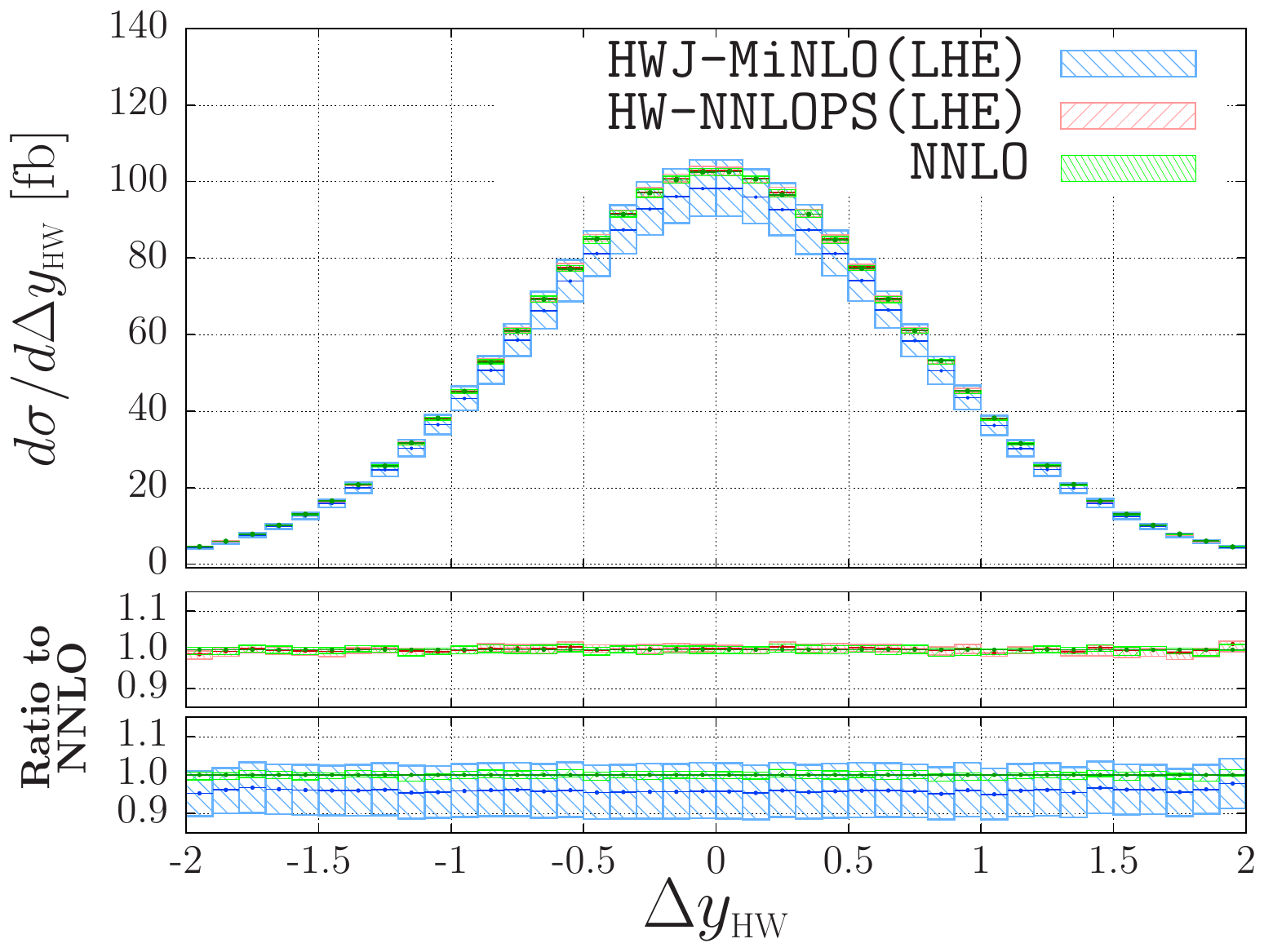} }}
  \caption{ Comparison of \HWJMINLOLHE{} (blue), 
    \HWNNLOLHE{} (red) and  \NNLO{} (green) predictions for the Born variables $\PhiHWsimp$ chosen to
    perform reweighting.  Left panels show rebinned distributions
    (used for reweighting), right panels show differential
    distributions with equispaced bins.  }
  \label{fig:born-check}
\end{figure}

For further validation we should check whether the distributions of
the Born variables $\PhiHWsimp$ obtained with {\HWNNLOLHE} reproduce
the results from the \HVNNLO{} code.  In Fig.~\ref{fig:born-check} we
can see rebinned distributions that we have used for reweighting
(left) and unrebinned distributions (right) of the rapidity of the
\HW{} system $\yhw{}$, the transverse momentum of Higgs boson $\pth{}$
and the rapidity difference between Higgs and W-boson $\Delta \yhw{}$.
We see that in the rebinned distributions we find perfect agreement
between \HWNNLOLHE{} and \NNLO{} results.
For the unrebinned distributions we see that, when rebinned bins are
large, e.g. for $|\yhw{}| \gtrsim 3$, minor artifacts
are present. These can be always reduced using a suitable, finer
binning for the 3D-histograms used for the reweighting.

As expected, the \HWNNLOLHE{} results reproduce very well results from
{\HVNNLO} and the uncertainty band of \HWJMINLOLHE{} shrinks from
around $\pm 10\%$ to about $\pm 2\%$ in the \HWNNLOLHE{} case, which
is a result of including NNLO corrections.

\subsection{Validation of the use of Collins-Soper angles}
\label{subsec:validation-cs-angles}

As discussed in the previous section the Collins-Soper (CS) frame is a
natural choice for the description of spin one vector boson decay.
This frame is convenient since it allows the angular dependence of the
vector decay to be parametrized in terms of only eight
coefficients. Here we want to verify how well the CS parametrization
works in practice.

In the case of $\thetacs$ distributions the only terms in
Eq.~\eqref{eq:sigma} that contribute are $A_{0}$ and $A_{4}$, since
the other terms drop out when integrating over $\phics$.  The $\phics$
distributions on the other hand depend only on $A_{2}, A_{3}, A_{5}$
and $A_{7}$.

\begin{figure}[htb]
  \centering
  {{ \includegraphics[width=0.48\textwidth]{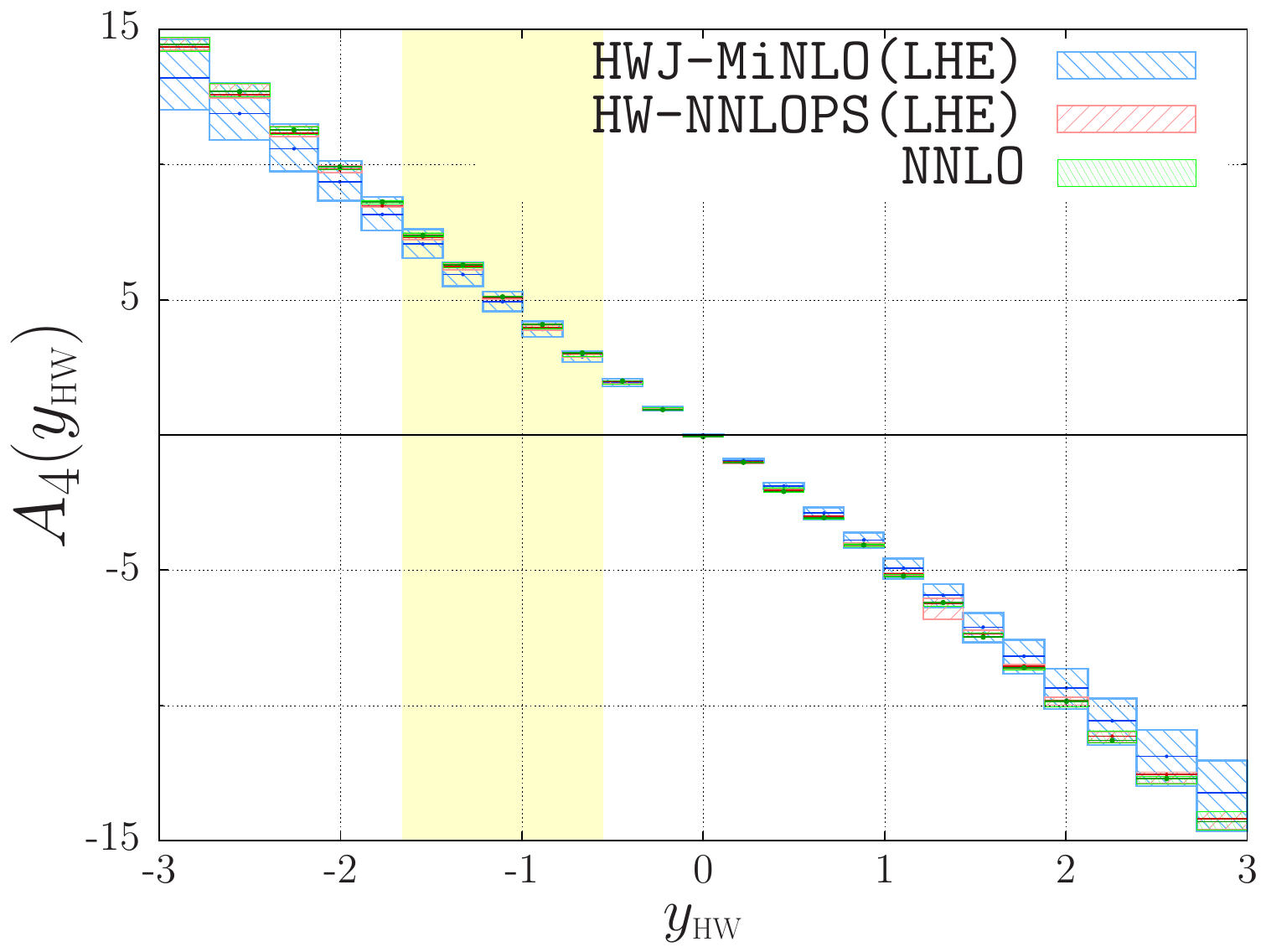} }}
  {{ \includegraphics[width=0.48\textwidth]{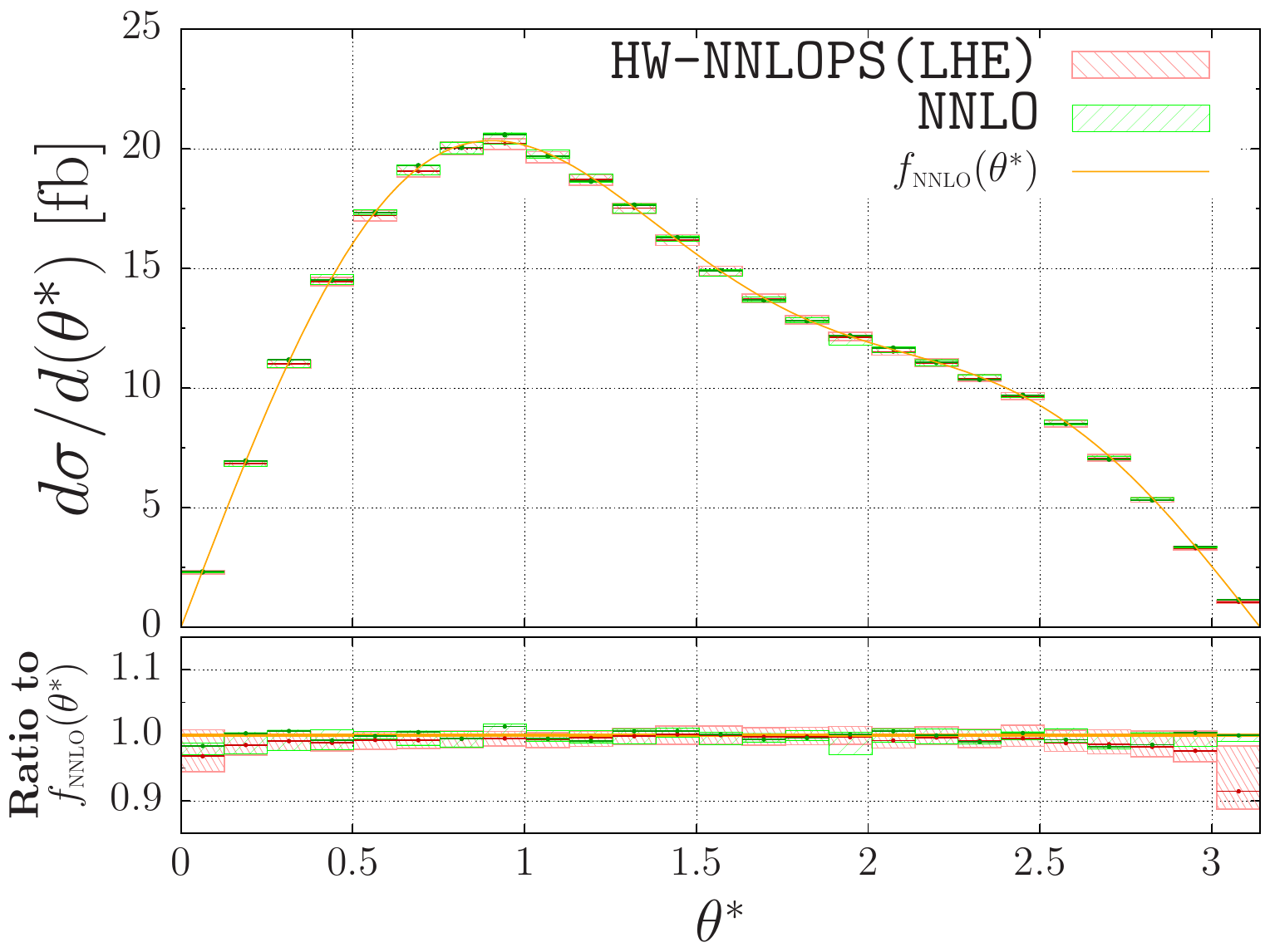} }}
  {{ \includegraphics[width=0.48\textwidth]{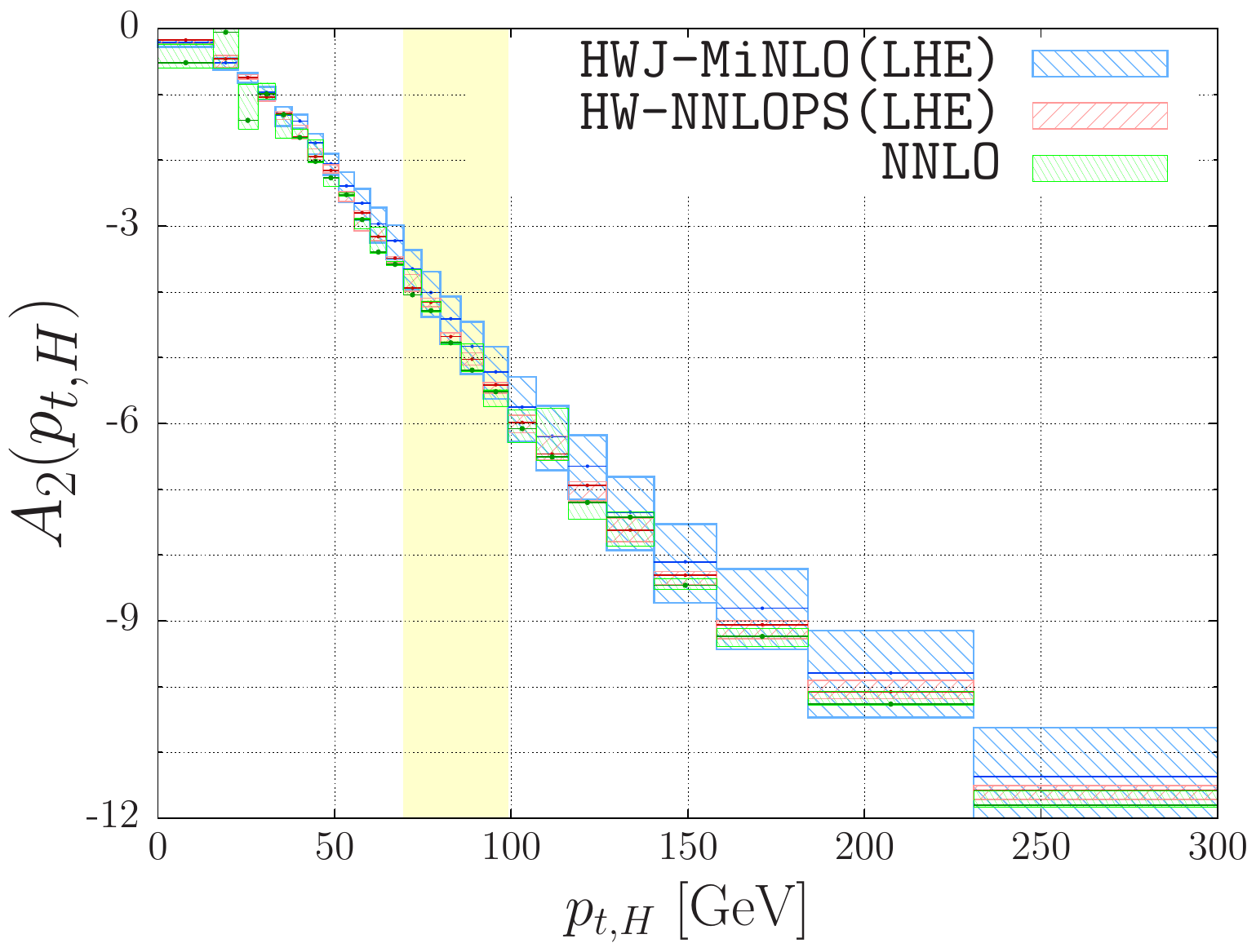} }}
  {{ \includegraphics[width=0.48\textwidth]{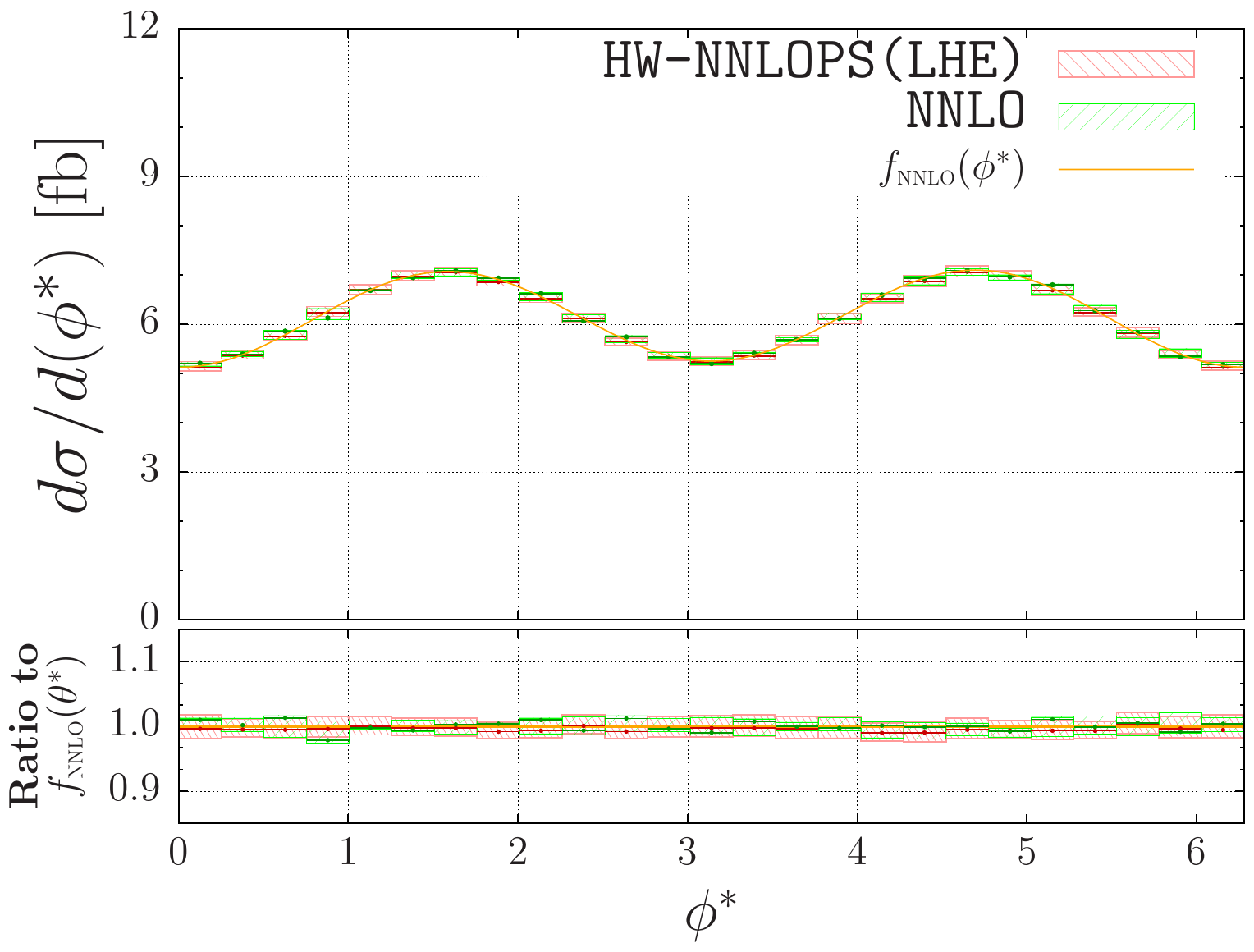} }}
  \caption{Upper panel: The left plot shows the $A_4$ coefficient as
    a function of $\yhw{}$. The right plot shows the distribution of
    $\thetacs$ integrated over all variables with $\yhw{}$ restricted
    to the region marked as yellow band in the left panel.  Lower
    panel: The left plot shows the $A_2$ coefficient as a function of
    $\pth{}$. The right plot shows the distribution of $\phics$
    integrated over all variables with $\pth{}$ restricted to the
    region marked as yellow band in the left panel.  }
  \label{fig:cs-plots}
\end{figure}

In the upper left panel of Fig.~\ref{fig:cs-plots} we show the
dependence of the coefficient $A_4$ on $\yhw{}$, whereas in the upper
right plot we present the $\thetacs$ distribution integrated over the
whole range of $\pth{}$, $\dyhw{}$ and, as an example, in the range of
$\yhw{}$ marked on the left upper plot by a yellow band. The red and
green bands denote the theoretical uncertainty, as described
before. The orange line shows the prediction from Eq.~\eqref{eq:sigma}
with the coefficients computed for the central scale choice from
Eq.~\eqref{eq:A} at pure NNLO level.  Notice that the $\thetacs$
distribution is not symmetric since we have restricted ourselves to
$\yhw$ values where $A_4$ is always positive, hence the functional
dependence encoded in $f_4(\thetacs,\phics)$ is visible.  From the
r.h.s. plot we can see that the central \NNLO{} result is fully
compatible with $f(\thetacs)$, i.e. the prediction from
Eq.~\eqref{eq:sigma}. Furthermore, we see that the \NNLO{} prediction
is consistent with the \HWNNLOLHE{} one, both for the central scale
and for the scale variation, as was the case for the other Born
variables used for reweighting.

Similar considerations apply to the $\phics{}$ dependence, whose shape
is determined by the $A_i$ coefficients, as the first term in
Eq.~\eqref{eq:sigma} integrates to a constant factor. We show in the
lower left panel of Fig.~\ref{fig:cs-plots} the dependence of the
coefficient $A_2$ on $\pth{}$ while integrated over the remaining
variables. In the lower right plot we display the distribution of
$\phics$ integrated over whole range of $\yhw{}$ and $\dyhw{}$, but
restricted to the $\pth{}$ interval highlighted with a yellow band in
the left plot.  As for the $\thetacs$ distribution, we have good
agreement between the \HWNNLOLHE{} result and the differential cross
section reconstructed from the CS parametrization. As expected, the
\NNLO{} prediction is also consistent with the \HWNNLOLHE{} one.
These and similar plots validate our use of the extraction of the
$A_i$ coefficients and their use to parametrize the angular
dependence.

\section{Phenomenological results}
\label{sec:pheno}

We will now discuss a few phenomenological results obtained with our
new code. We remind the reader that the specific process studied here
is $ pp\rightarrow H\ell^{+}\nu_{\sss\ell}$, with $\ell^+ = \{e^+,
\mu^+\}$ and that we do not consider decays of the Higgs boson.

For all the results presented in this section we apply the cuts that
were suggested in the context of the Higgs Cross Section Working Group
(HXSWG) activity for the preparation of the fourth Yellow Report.  We
consider $13$~TeV LHC collisions. We require one positively charged
lepton with $|y_\ell |< 2.5$ and $p_{t,\ell} > 15$ GeV, while we do
not impose a missing energy cut. When applying a jet-cut or a jet-veto
we define a jet as having $p_{t,j} > 20$ GeV and $|y_{j}| < 4.5$. Jets
are reconstructed using the anti-$k_t$
algorithm~\cite{Cacciari:2008gp} with $R=0.4$, as implemented in
\texttt{Fastjet}~\cite{Cacciari:2011ma}. At the moment we do not apply
any cuts on the Higgs boson, however our code produces Les Houches
events, which can be interfaced with any tool that provides the decay
of the Higgs in the narrow width approximation. For example, this can
be obtained easily by allowing \PYTHIA{8} to treat the Higgs boson as
an unstable object.

\subsection{Fiducial cross-section}

The fiducial cross section at $\sqrt{s}=13$\,TeV, together with its
theoretical uncertainty, at different levels of the simulation, is
given in table~\ref{tab:total-xs}.
From these results we obtain a $K$-factor between {\HVNNLOPS} and
{\HWJMINLO} equal to $1.04$.
\begin{table}[h]
  \centering
  \begin{tabular}{|c|c|c|c|}
    \hline
    & {\HWJMINLO} & {\HVNNLO} & {\HVNNLOPS}  \\
    \hline
    $\sigma_{tot}$
    & $\quad 152.49(5)\; $fb$ \> \pm 7.0\% \quad$    
    & $\quad 158.75(8)\; $fb$ \> \pm 1.0\% \quad$
    & $\quad 159.21 (30)\; $fb$ \> \pm 1.0\% \quad$ \\
    \hline    
  \end{tabular}
  \caption{Fiducial cross-section of $pp\rightarrow HW^{+}
    \rightarrow H\ell^{+}\nu_{\ell}$ at $\sqrt{s}=13$ TeV with
    leptonic cuts.  The uncertainty band is obtained with the scale
    variation procedure described in the text. Numerical errors for
    each prediction are quoted in brackets, and relative details are
    given in the text.}
  \label{tab:total-xs}
\end{table}
We also see that the reweighting procedure of {\HWJMINLO} events to
\NNLOPS{} accuracy gives a result compatible with the fixed order NNLO
calculation. In particular, the sizes of scale uncertainties for the
\HVNNLOPS{} and {\HVNNLO} results are fully comparable, providing a
reduction of almost one order of magnitude with respect to the
\HWJMINLO{} result. The number quoted in bracket for the \HWJMINLO{}
and \HVNNLO{} results is the statistical error, and it is entirely due
to Monte Carlo integration. The \HVNNLOPS{} statistical uncertainty
was found to be compatible with the one of \HWJMINLO{}. The numerical
error quoted for the \HVNNLOPS{} result is larger because it also
contains a systematic component, that we added in quadrature to the
statistical one, and which is due to bin-size effects in the reweighting
procedure.\footnote{This error has been estimated by varying the
  number of bins in the reweighting procedure described in
  Sec.~\ref{sec:practical}, and also by performing a reweighting
  without taking into account the dependence on the Collins-Soper
  angles.}
%

\subsection{Higgs and Leptonic Observables}

In the following we consider cross-sections obtained at various
levels: at Les Houches event level before shower at NLO or NNLO
accuracy, \HWJMINLOLHE{} and \HWNNLOLHE{}, respectively; after
showering the \HWJMINLOLHE{} and \HWNNLOLHE{} events with \PYTHIA{8},
\HWJMINLOPS{} and \HWNNLOPSpyth{}, both with and without
hadronization.

\begin{figure}[!htb]
  \centering
  \includegraphics[width=0.48\textwidth]{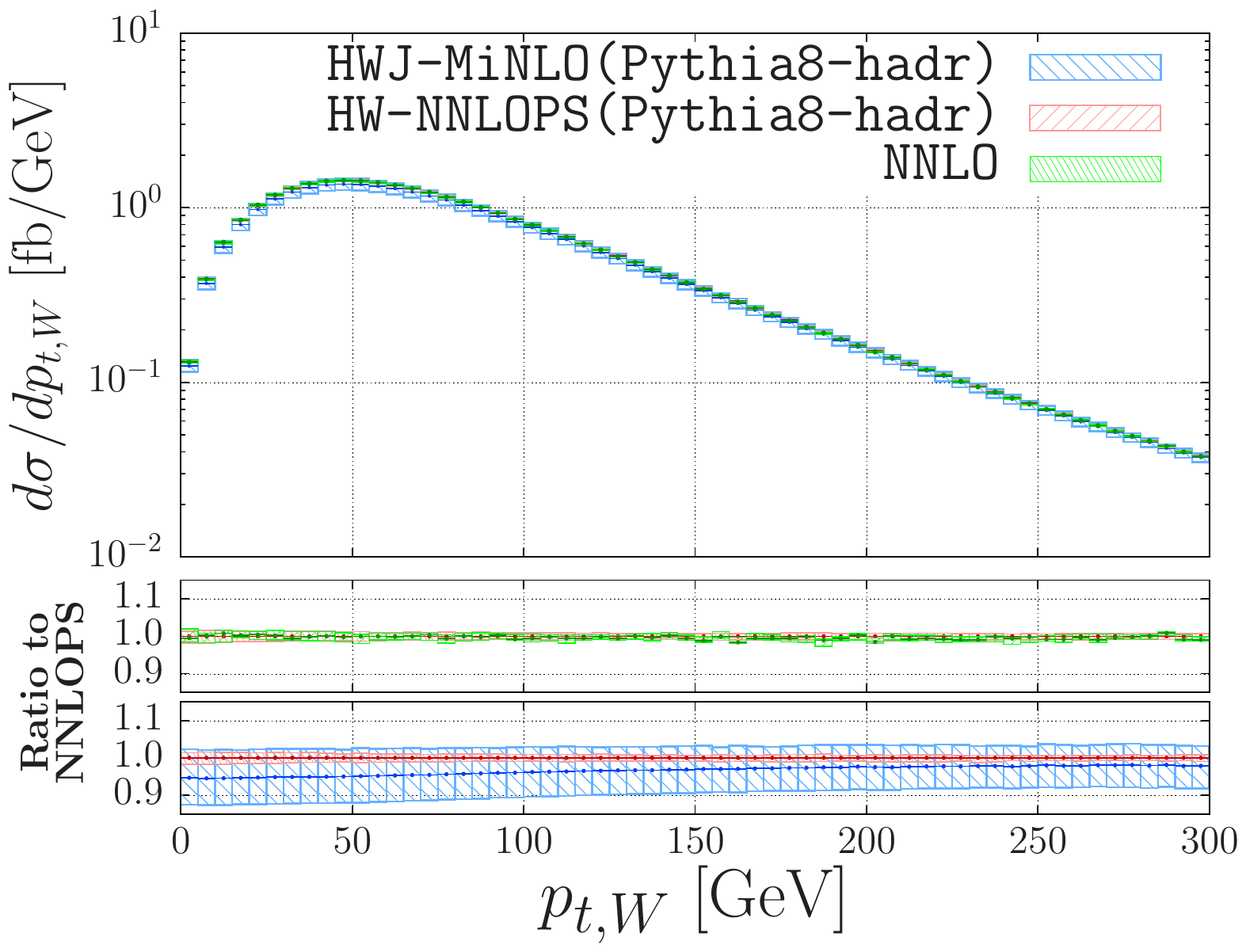}
  \includegraphics[width=0.48\textwidth]{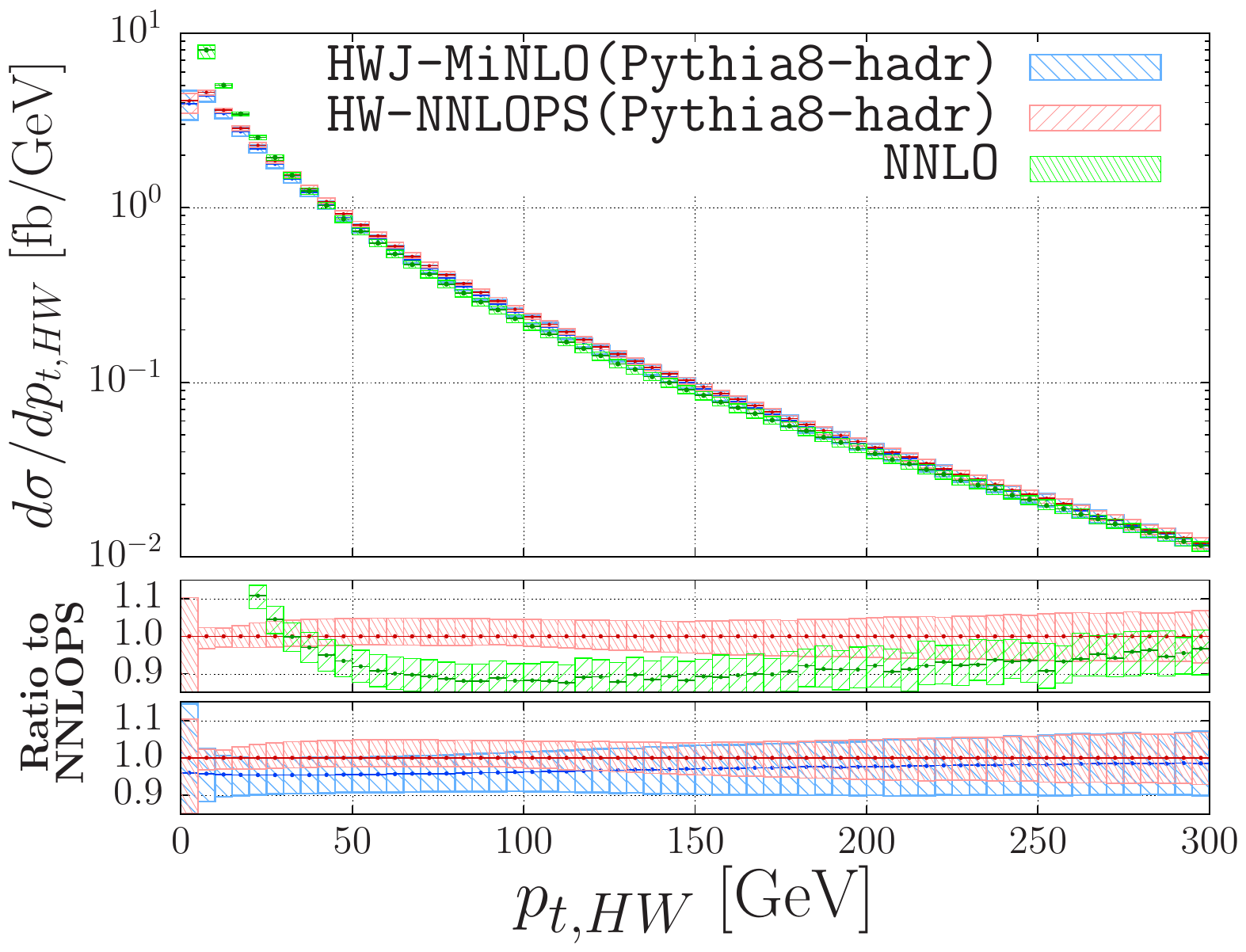}
  \caption{Comparison of \HWJMINLO{} (blue), NNLO (green), and \HWNNLOPS{} (red) predictions for $\ptw$ (left) and $\ptwh$ (right). }
  \label{fig:extra_ptw_ptwh}
\end{figure}

We start by showing in Fig.~\ref{fig:extra_ptw_ptwh} the distributions
for the transverse momenta of the $W$ boson and the \HW{} system,
respectively.
NNLO results are compared against those obtained with \HWJMINLO{} and
\HVNNLOPS{}. For observables that are fully inclusive over QCD
radiation, such as $\ptw$, the agreement among the \HVNNLO{} and
\NNLOPS{} predictions is perfect, as expected.  As in the case of the
fiducial cross-section one notices the sizable reduction of the
uncertainty band from around $7$\% in \HWJMINLO{} to about $1\%$ in
the case of \HVNNLO{} and \HVNNLOPS{}.
As no particularly tight cuts are imposed, the NNLO/NLO $K$-factor is
almost exactly flat.

The right panel shows instead the effects due to the Sudakov
resummation. At small transverse momenta, the NNLO cross section
becomes larger and larger due to the singular behaviour of the matrix
elements for \HW{} production in association with arbitrarily
soft-collinear emissions. The \MINLO{} method resums the logarithms
associated to these emissions, thereby producing the typical Sudakov
peak, which for this process is located at $2\text{ GeV
}\nobreak\lesssim\nobreak\ptwh \lesssim 5\text{ GeV}$, as expected
from the fact that the LO process is Drell-Yan like.  It is also
interesting to notice here two other features that occur away from the
collinear singularity, and which are useful to understand the plots
which are shown later. Firstly, the $\pt$-dependence of the NNLO
reweighting can be explicitly seen in the bottom panel, where one can
also appreciate that at very large values not only the \NNLOPS{} and
\MINLO{} results approach each other, but also that the uncertainty
band of \HVNNLOPS{} becomes progressively larger (in fact, in this
region, the nominal accuracy is NLO). Secondly, in the region
$30\text{ GeV}\nobreak\lesssim\nobreak\ptwh\nobreak \lesssim\nobreak
250\nobreak\text{ GeV}$, the NNLO and \NNLOPS{} lines show deviations
of up to $10$\,\%: these are due to both the compensation that needs
to take place in order for the two results to integrate to the same
total cross section, and the fact that the scale choices are different
(fixed for the NNLO line, dynamic and set to $\ptwh$ in
\MINLO{}). When $\ptwh\gtrsim 250$ GeV the two predictions start to
approach, as this is the region of phase space where the \MINLO{}
scale is similar to that used at NNLO ($\mu = M_{\scriptscriptstyle
  H}+M_{\scriptscriptstyle W}$).  At even higher transverse momenta,
$\ptwh \gtrsim 400$ GeV, the \MINLO{} Sudakov is not active, however
the \MINLO{} scale is set to the transverse momentum which is higher
than the scale in the NNLO calculation. As consequence, the \NNLOPS{}
results are lower than the NNLO one.

\begin{figure}[h]
  \centering
  {{ \includegraphics[width=0.48\textwidth]{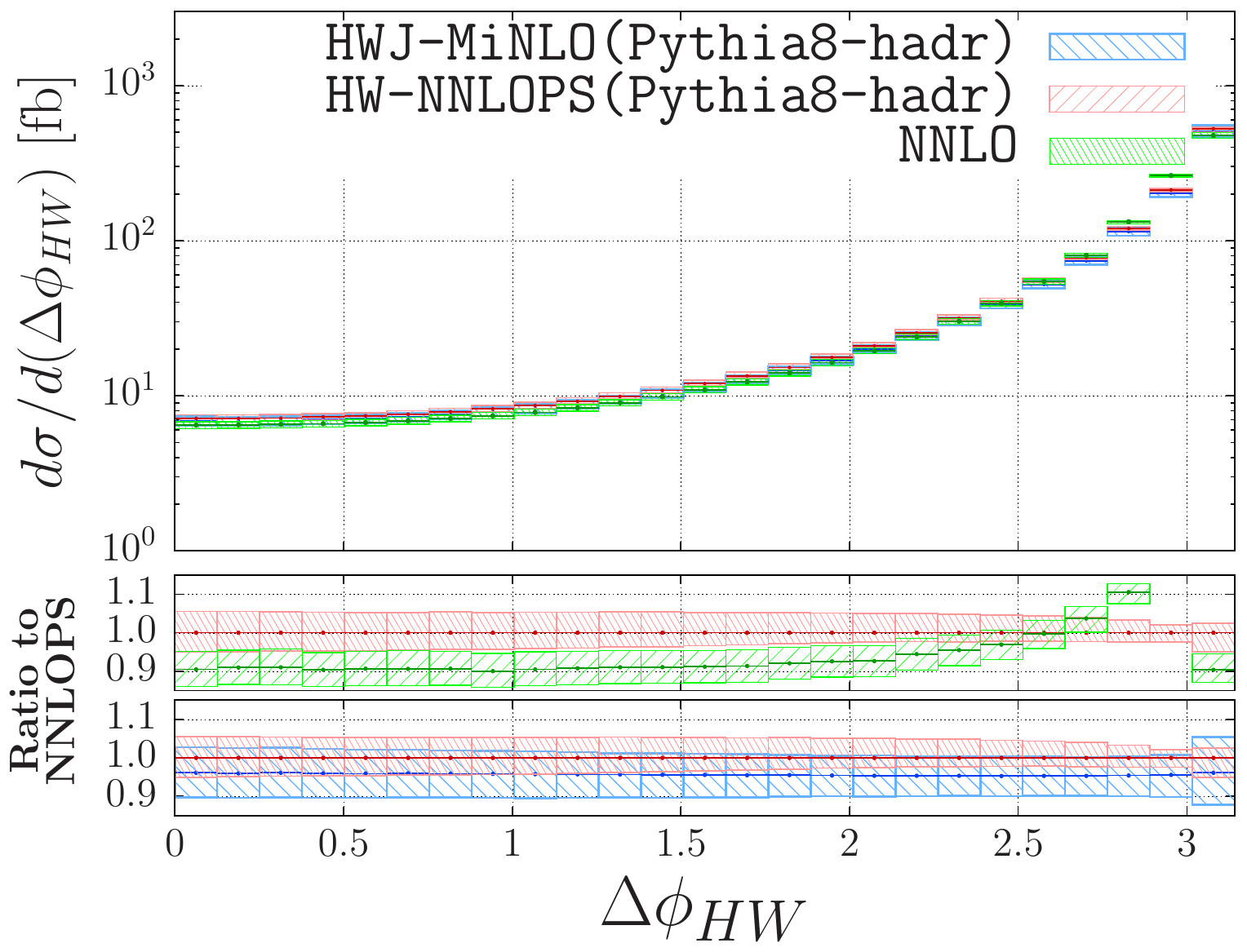} }}
  {{ \includegraphics[width=0.48\textwidth]{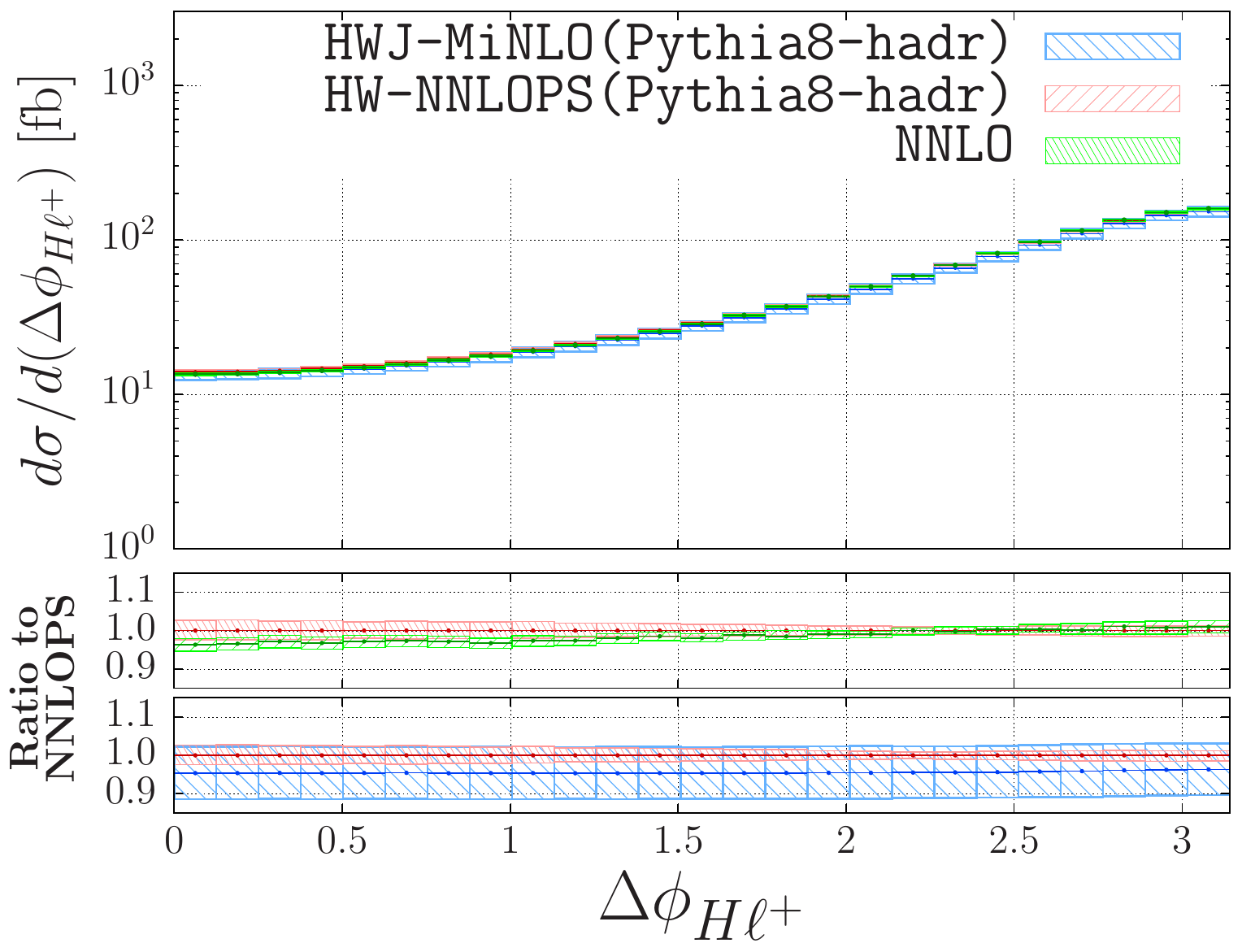} }}
  \caption{Azimuthal angle between the Higgs boson and the $W^{+}$ boson
    ($\Delta\phi_{\sss \HW{}}$, left) and azimuthal angle between the Higgs boson
    and the charged lepton ($\Delta\phi_{\sss H\ell^{+}}$, right).  }
  \label{fig:VH-delphi}
\end{figure}

It is interesting to look at a variable describing the decay of the
\HW{} resonance, e.g. the azimuthal angle between the $W^{+}$ boson
and the Higgs particle ($\Delta\phi_{\HW}$).  At leading order the two
particles are back-to-back, $\Delta\phi_{\HW}=\pi$, but real radiation
moves the bosons away from this configuration.
In Fig.~\ref{fig:VH-delphi} (left) we show the distribution of
$\Delta\phi_{\HW{}}$ comparing the \HVNNLO{} result to the result of
our simulation after including parton shower effects, before and after
the NNLO rescaling.
For moderate values of $\Delta\phi_{\HW}$ ($\lesssim 2.0$) we have a
very flat correction, as this region is dominated by events with high
transverse momentum of the \HW{}-system, and dominant effects captured
by fixed order NNLO calculation.  However, the limit with nearly
back-to-back emission of $H$ and $W^{+}$ corresponds to the
low-$p_{t,\HW{}}$ region which is sensitive to the effects of soft
radiation.  Hence there are pronounced differences in the region
$\Delta\phi_{\HW} \gtrsim 2.5 $ between the \NNLOPS{} simulation, and
the NNLO prediction that diverges at $\Delta\phi_{\HW}=\pi$.
On the contrary, the distribution of the azimuthal angle between
$\ell^{+}$ and Higgs, shown in the right panel of
Fig.~\ref{fig:VH-delphi}, has no divergence in the NNLO
calculation. It therefore has a much flatter $K$-factor throughout the
whole range, and the theoretical uncertainty bands of the \HVNNLO{}
and \HVNNLOPS{} simulations mostly overlap.

\begin{figure}[h]
  \centering
  {{ \includegraphics[width=0.48\textwidth]{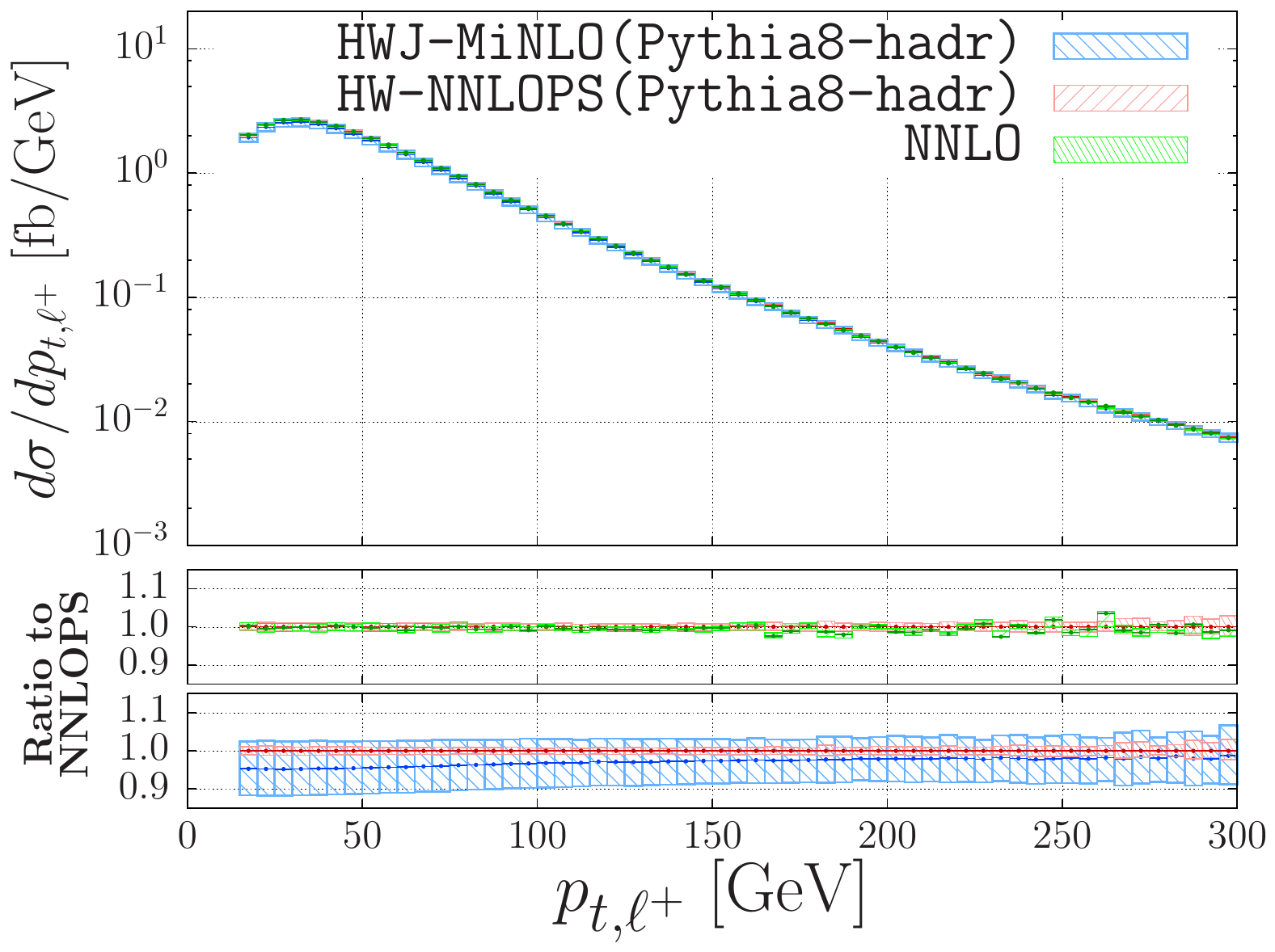} }}
  {{ \includegraphics[width=0.48\textwidth]{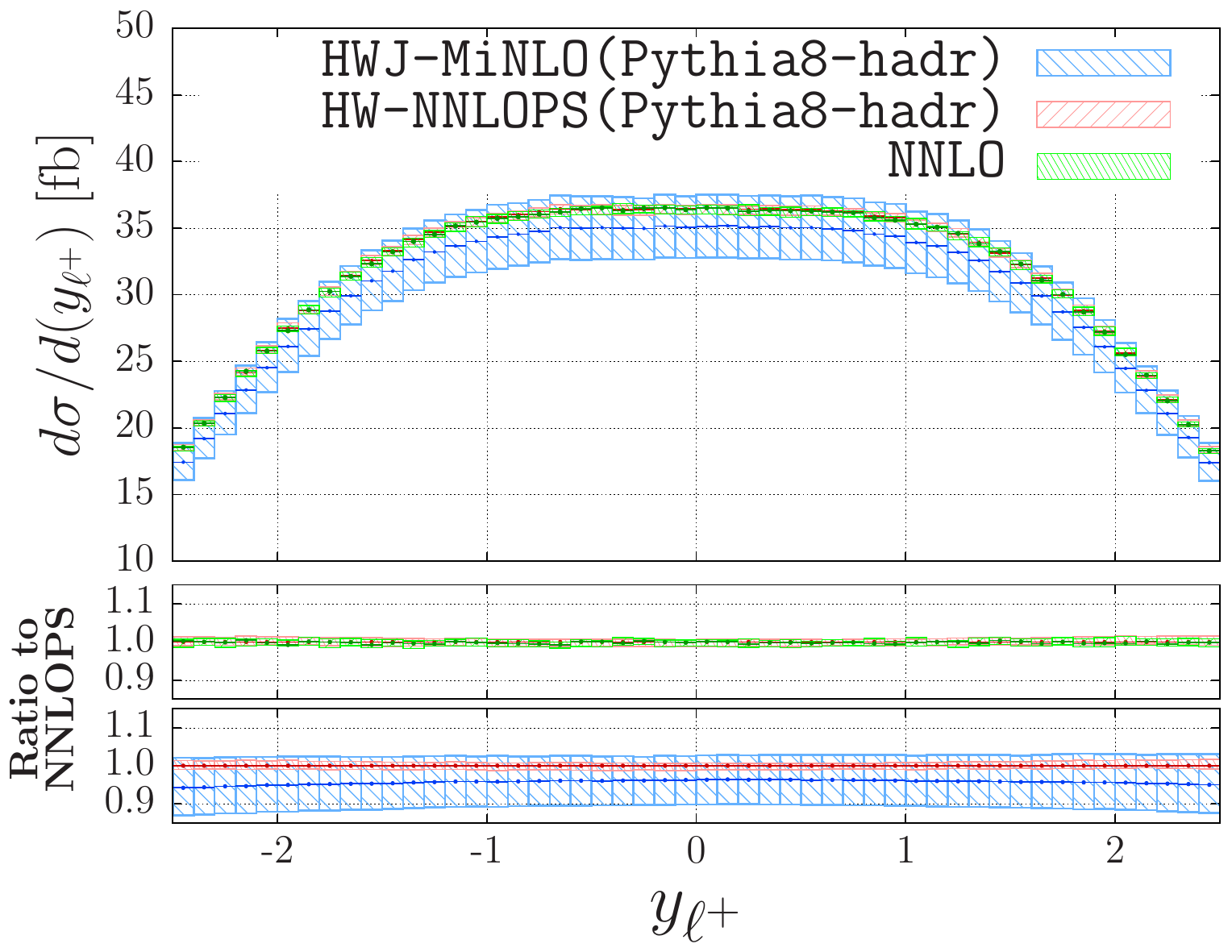} }}
  \caption{
    Transverse momentum and rapidity of the positively charged lepton $\ell^{+}$.
}
  \label{fig:l1-pt-y}
\end{figure}

\begin{figure}[h]
  \centering
  \includegraphics[width=0.48\textwidth]{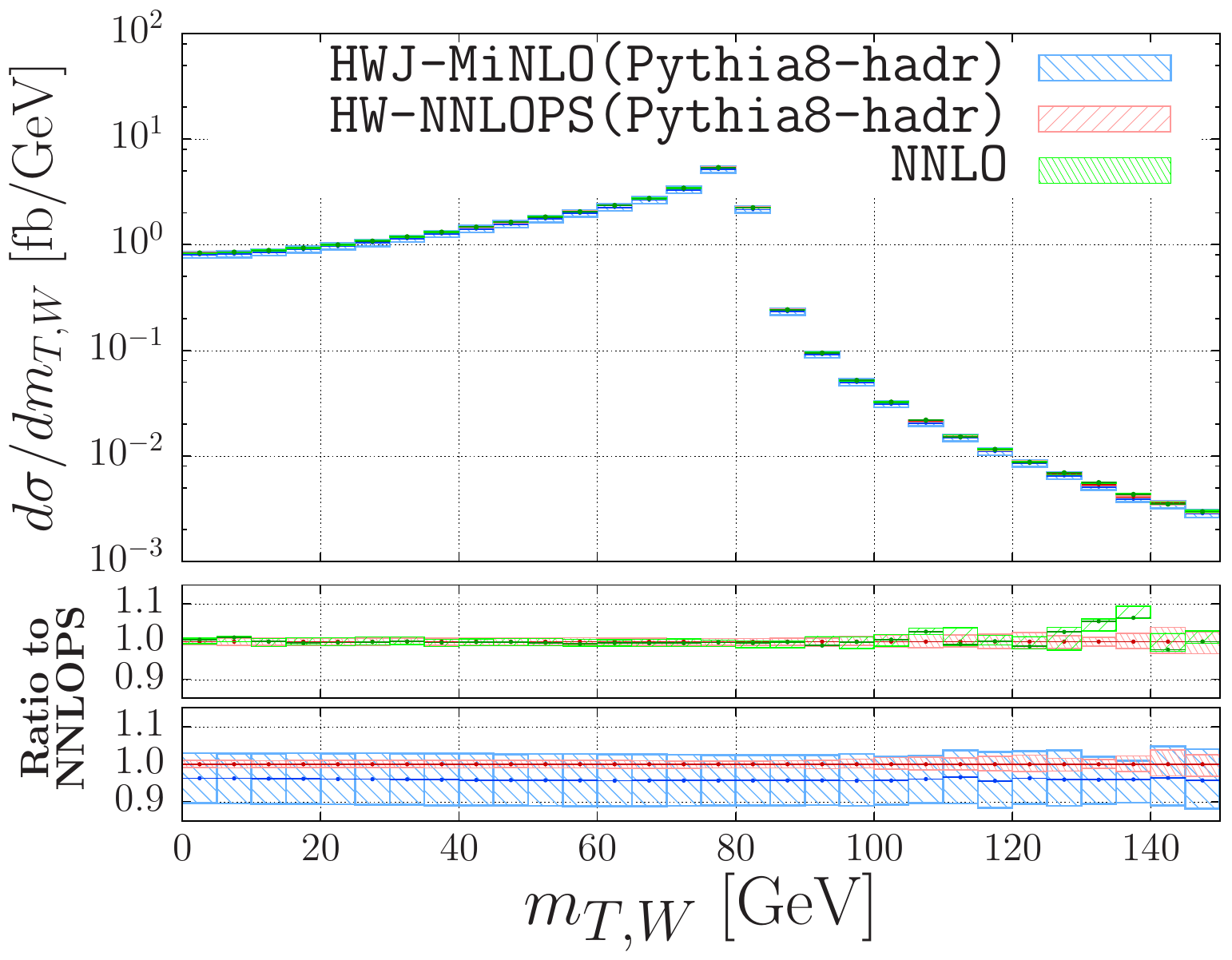}
  \includegraphics[width=0.48\textwidth]{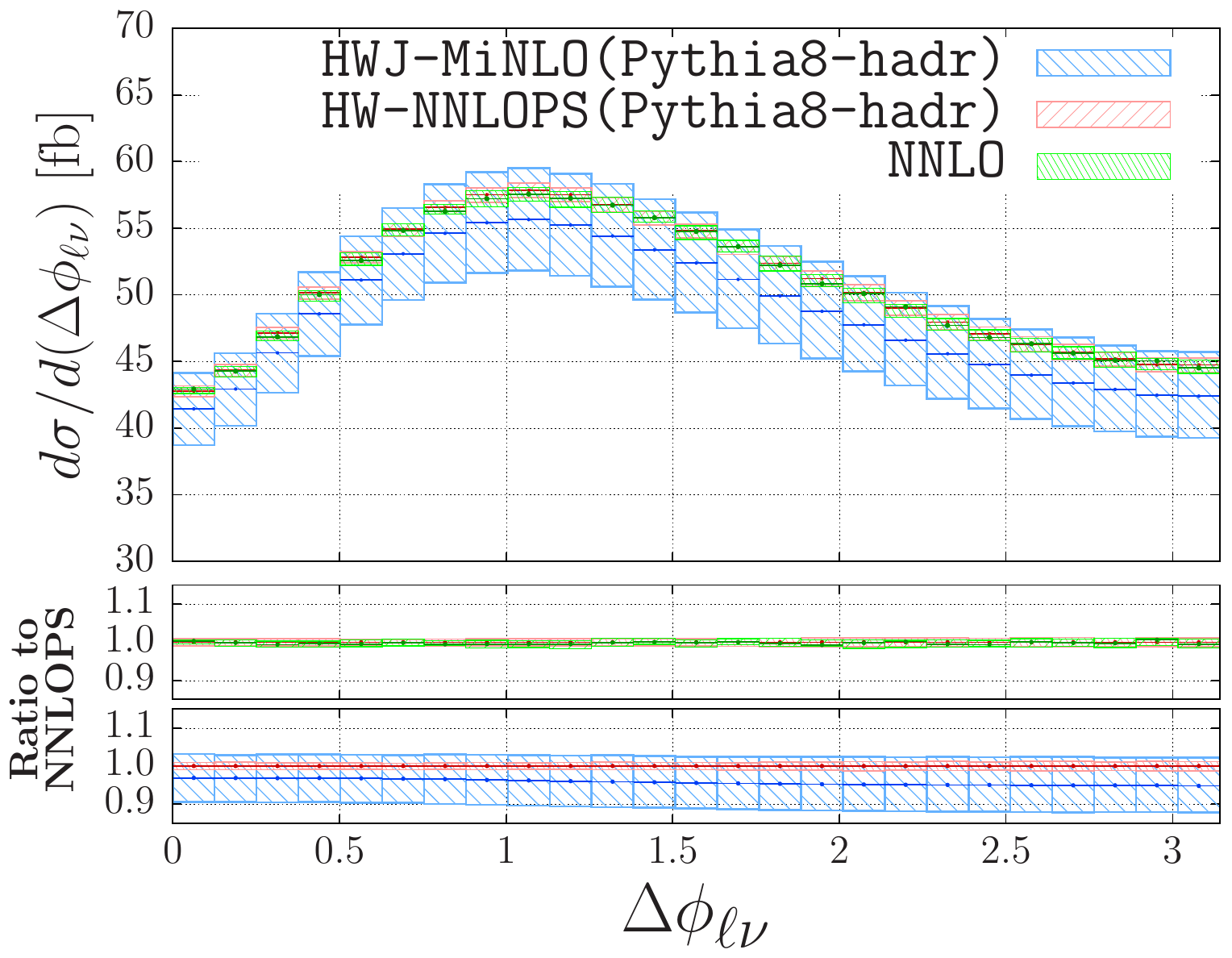}
  \caption{The transverse mass of the $W^{+}$ boson (left) and the
    azimuthal angle between $\ell^{+}$ and the neutrino
    (right). }
  \label{fig:mT-l1l2-dphi}
\end{figure}

We next present in Fig.~\ref{fig:l1-pt-y} the distributions of the
transverse momentum (left) and the rapidity (right) of the positive
lepton $\ell^{+}$.
We can see that there is a clear agreement between NNLO predictions
and \NNLOPS{} results.  
Other interesting variables are the azimuthal angle between $\ell^{+}$
and the neutrino, $\Delta
\phi_{\sss\ell^{+}\nu}$, and the transverse mass of the $W^{+}$ boson,
defined as
\begin{equation}
  m_{\sss T,W} = \sqrt{2 p_{t,\nu}\,p_{t,\ell+} ( 1 - \cos(\Delta \phi_{\ell+,\nu}) )}\,.
\end{equation}
These two variables have characteristic shapes and we show in
Fig.~\ref{fig:mT-l1l2-dphi} that, as expected, our \NNLOPS{} code
agrees very well with pure NNLO predictions.

\subsection{Jet observables}
\label{subsec:jet-pheno}

We present now the study of observables involving final state jets. We
will focus on the differences in distributions coming from NNLO, and
\HVNNLOPS{} at both parton and hadron level.
\begin{figure}[h]
  \centering
  {{ \includegraphics[width=0.48\textwidth]{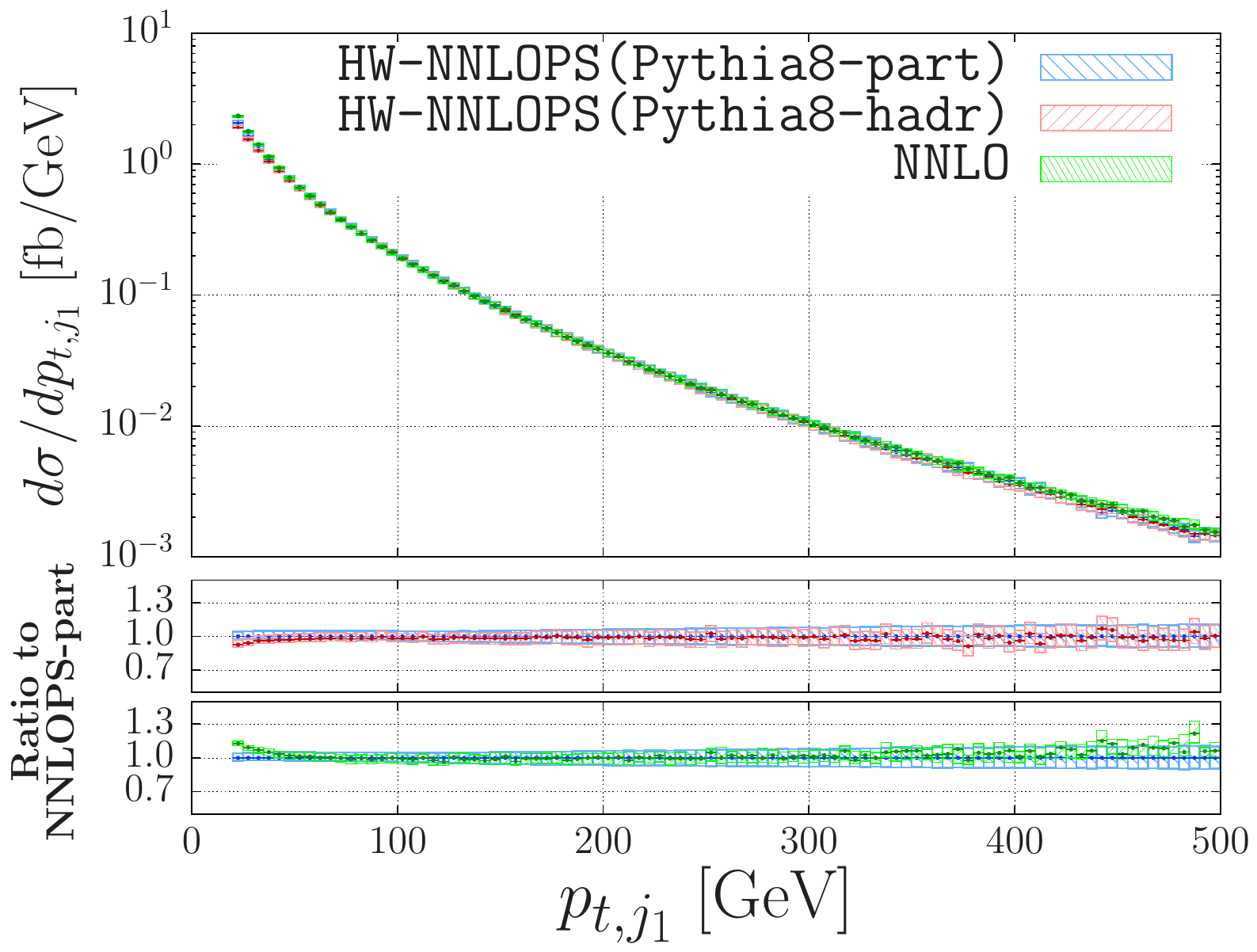} }}
  {{ \includegraphics[width=0.48\textwidth]{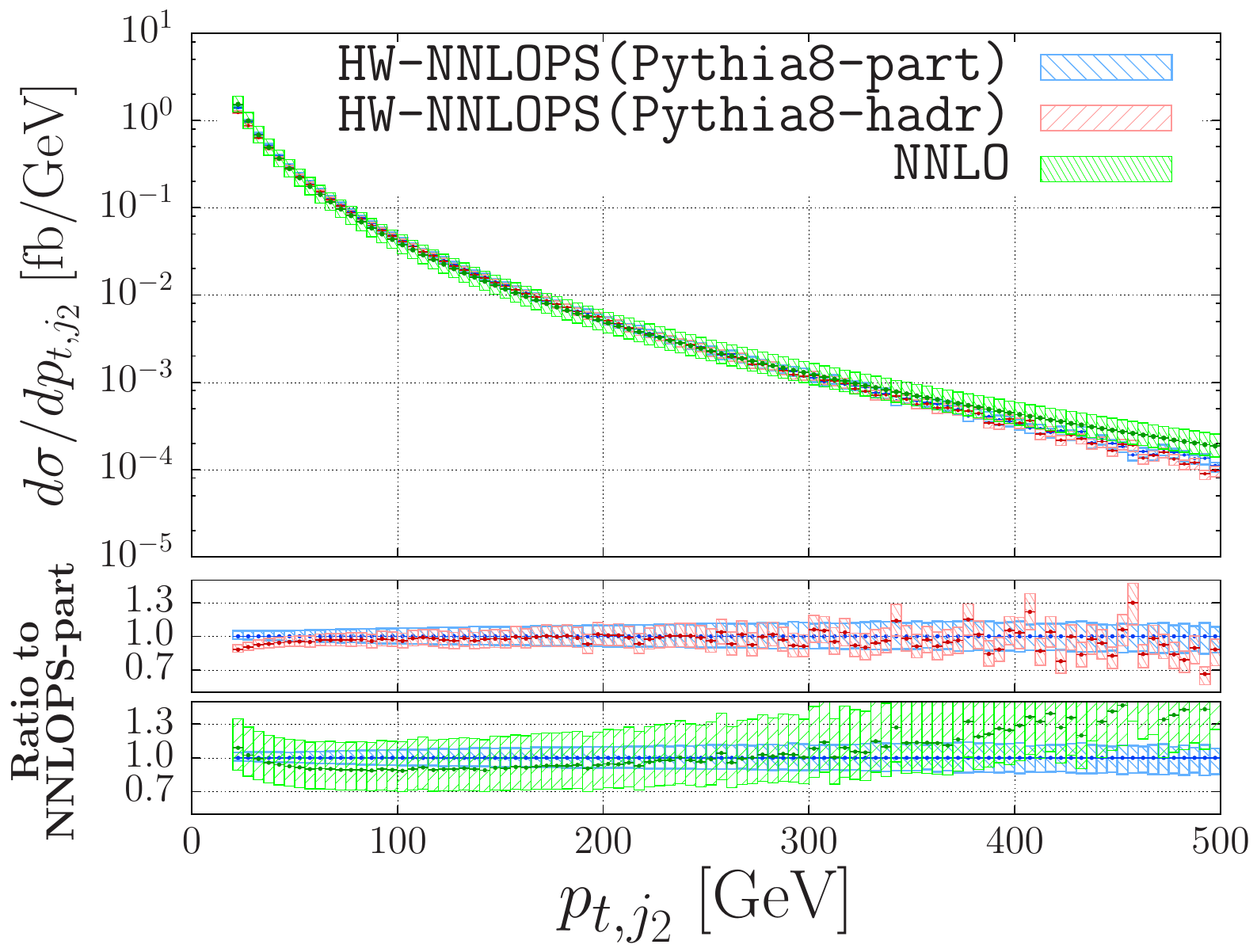} }}
  \caption{The transverse momentum of the two hardest jets at
  \NNLO{} (green), \HWNNLOPS{} before hadronization (blue) and \HWNNLOPS{} with hadronization (red).}
  \label{fig:pt-j1j2}
\end{figure}
In Fig.~\ref{fig:pt-j1j2} we show the transverse momentum of two
hardest jets. The distributions are cut at the minimum transverse
momentum used for jets, i.e. $20$\,GeV, but, from the ratio plot, one can
see that the fixed order NNLO calculation starts increasing sharply as
it approaches a divergence at low-$p_{t}$. As we identify only jets
with $p_{t}>20$~GeV we do not see the Sudakov peak in the
\HWNNLOPS{} simulations, which sits below the cut.

We will first discuss differences between the pure fixed order
calculation (green) and the \NNLOPS{} result before hadronization
(blue).
At large transverse momenta, theoretical uncertainties for the first
jet (Fig.~\ref{fig:pt-j1j2}, left) are of comparable size in all
simulations, even if they are slightly smaller in the NNLO
calculation. We should also note that, as in the case of
$\ptwh$, the \HVNNLO{} result is larger than the \HVNNLOPS{} one for
large-$p_{t}$ values.  This behaviour is a result of using a fixed
scale in the former, and a dynamical scale in the latter code.

For the second jet transverse momentum distribution
(Fig.~\ref{fig:pt-j1j2}, right), we note that, as expected, the
theoretical uncertainty is larger than in the previous case, as the
second jet is described only with LO accuracy. However we note that
the scale variation procedure now gives smaller bands for the
\HVNNLOPS{} simulation, compared to the NNLO calculation.  This is due
to the fact~\cite{Hamilton:2013fea} that \POWHEG{} produces additional
radiation (the second jet in the case of \HWJMINLO{}) with a procedure
that is insensitive to scale variation. The second jet spectrum is
multiplied by the NLO cross section kept differential only in the
underlying Born variables, i.e. the $\bar{B}$ function. Scale
variation affects only the computation of this function (which is NLO
accurate), hence as a result the uncertainty due to scale variation
for the $\ptjtwo$ spectrum is underestimated with respect to a
standard fixed-order computation. We recall that this is a known issue
in \POWHEG{} simulations, and was discussed in several previous
publications~\cite{Alioli:2008tz,Alioli:2010xd}. In order to get a
more reliable uncertainty band, one can split the real contribution
into a singular part (which enters in both the $\bar{B}$ function and
the \POWHEG{} Sudakov) and a finite one, corresponding to two resolved
emissions. By not including the latter contribution in the $\bar{B}$
function and in the \POWHEG{} Sudakov, the estimation of scale
uncertainty would be more similar to what one expects for an
observable which is described at LO, as the second-jet high-$p_T$
tail.

\begin{figure}[h]
  \centering
  {{ \includegraphics[width=0.48\textwidth]{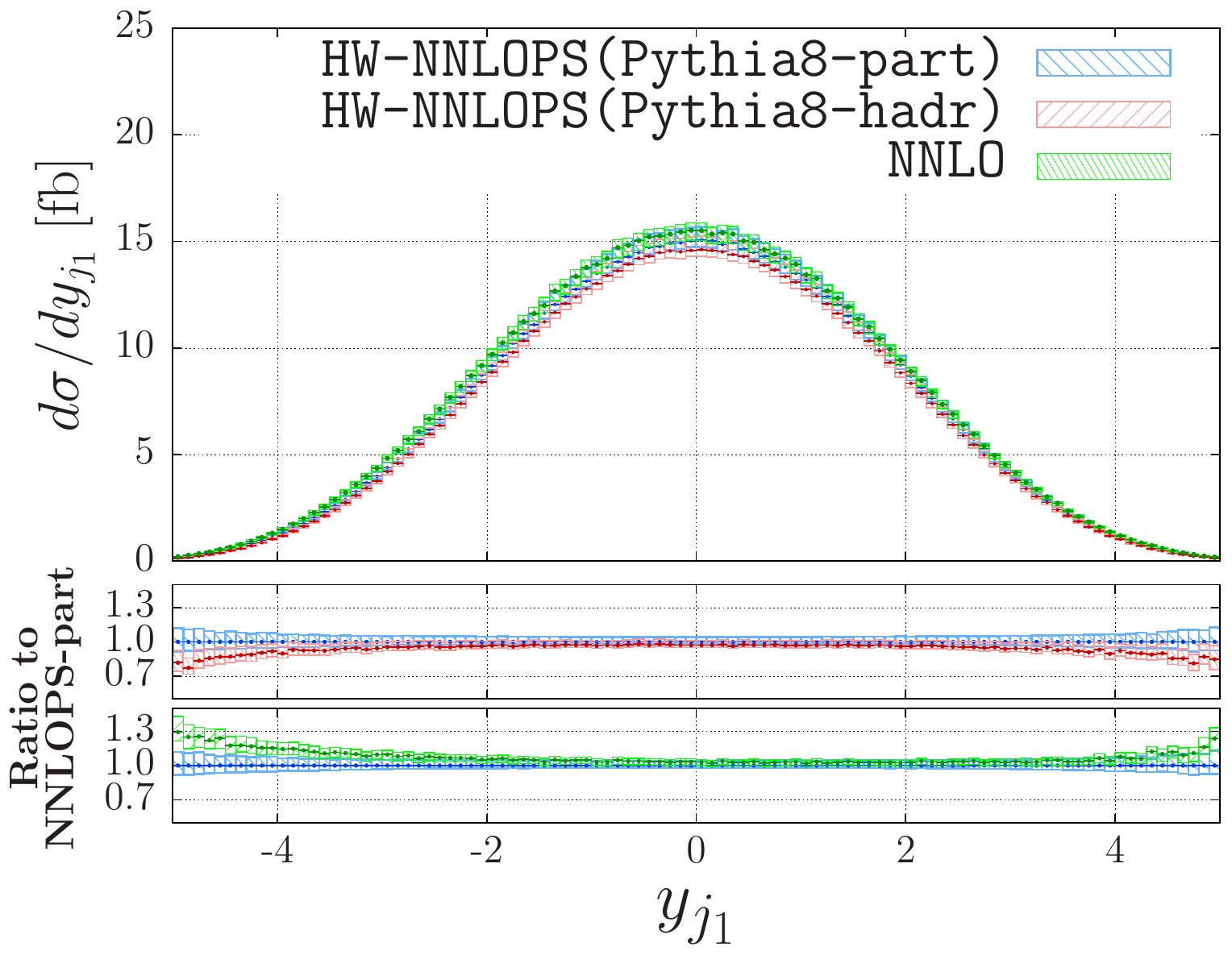} }}
  {{ \includegraphics[width=0.48\textwidth]{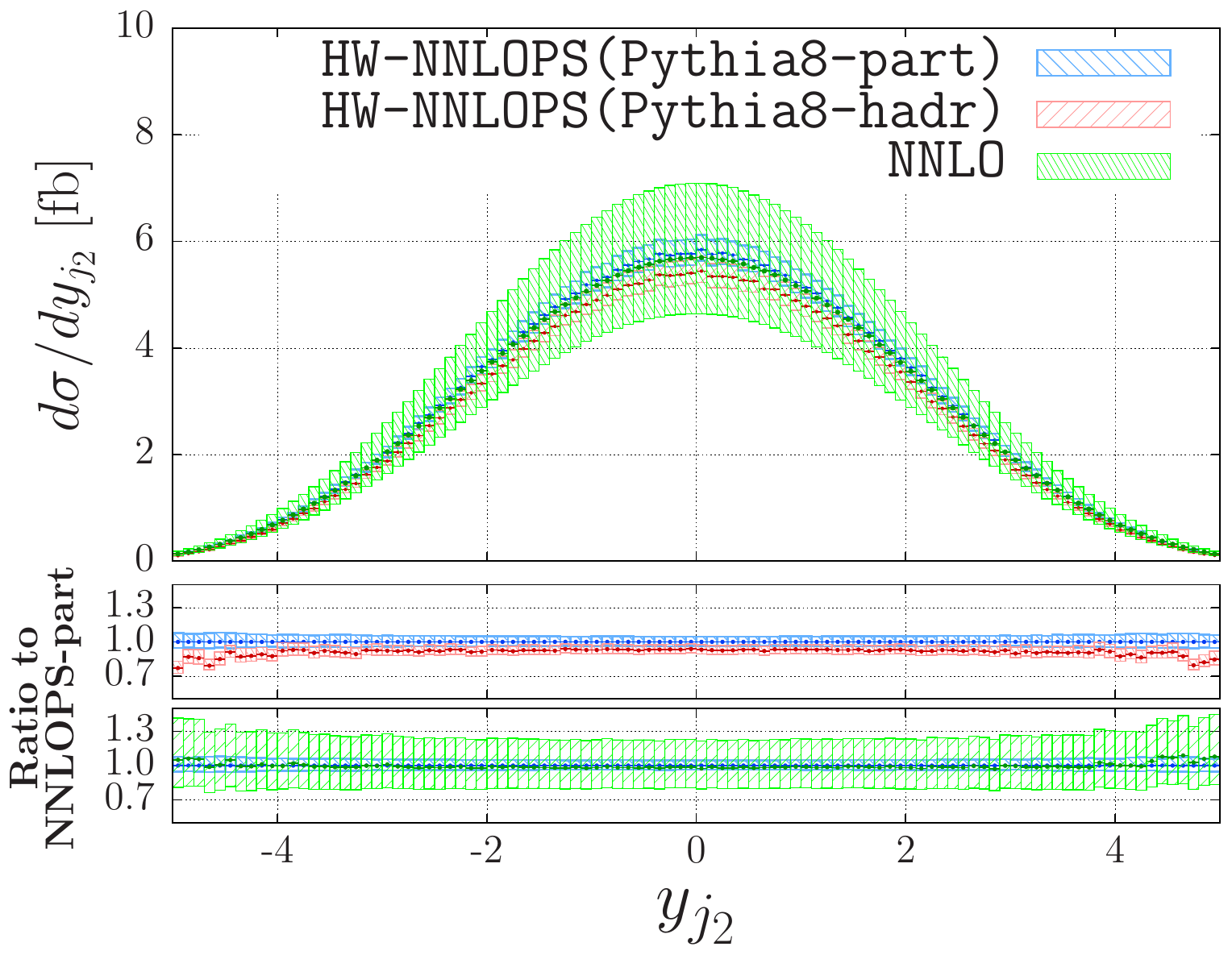} }}
  \caption{The rapidity of the two hardest jets 
 in \NNLO{} (green), \HWNNLOPS{} before hadronization (blue) and \HWNNLOPS{} with hadronization (red). }
  \label{fig:eta-j1j2}
\end{figure}

Next we find it interesting to examine the size of non-perturbative
effects.  Hadronization has a sizable impact on the shapes of jet
distributions: differences up to
$7\hspace{-0.05cm}-\hspace{-0.05cm}8$\,\% can be seen in the $p_{t,j1}$
spectrum at small values, and are still visible at a few percent level
till  relatively hard jets are required ($\ptjone > 100$ GeV). For
the second jet, hadronization corrections are similar and only
slightly more pronounced.
Even larger effects can be seen in the rapidity distribution of the
two leading jets at large rapidities, as can be seen from
Fig.~\ref{fig:eta-j1j2}. This is not surprising since the large
rapidity region is dominated by small transverse momenta.

\begin{figure}[h]
  \centering
  \includegraphics[width=0.48\textwidth]{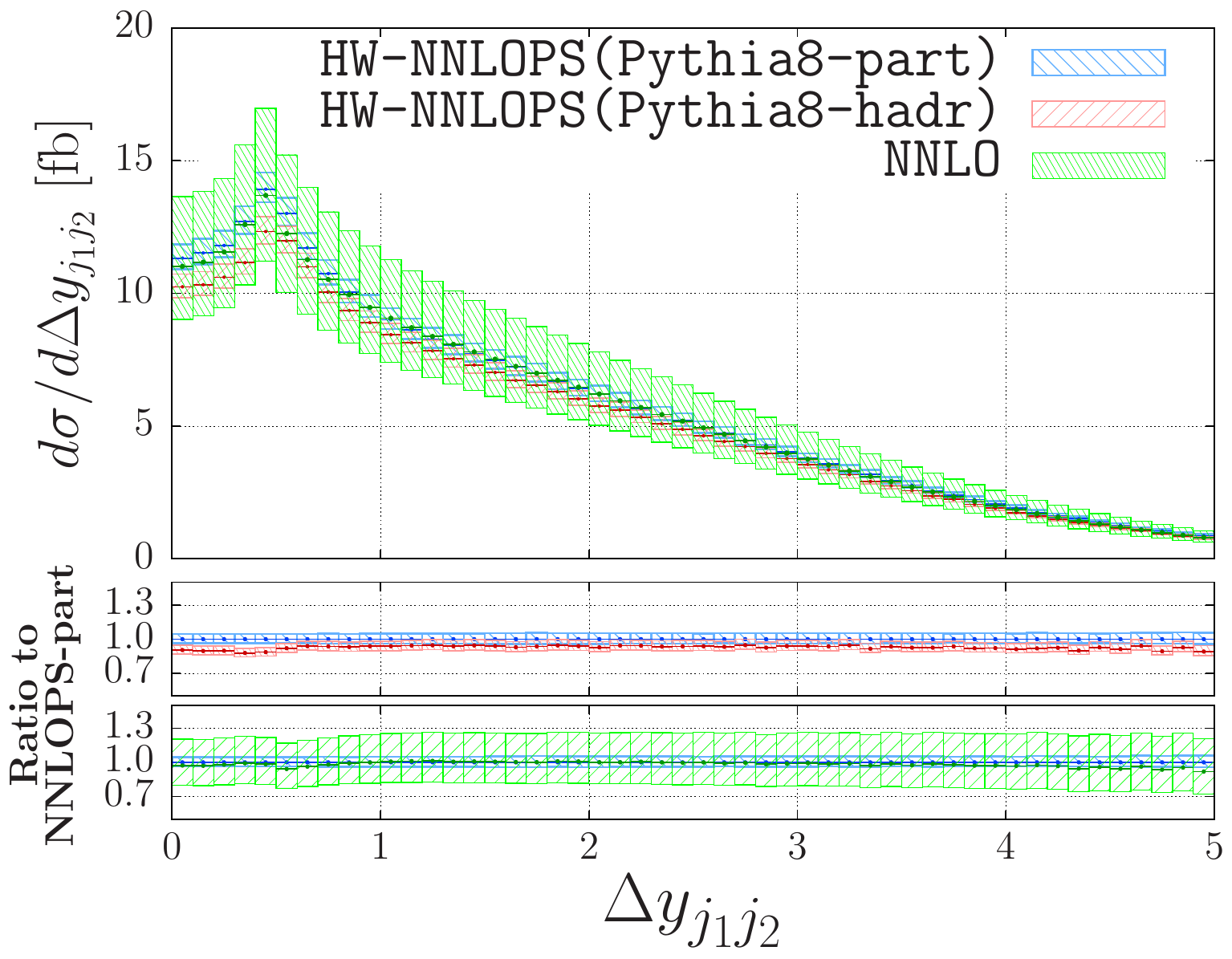}
  \includegraphics[width=0.48\textwidth]{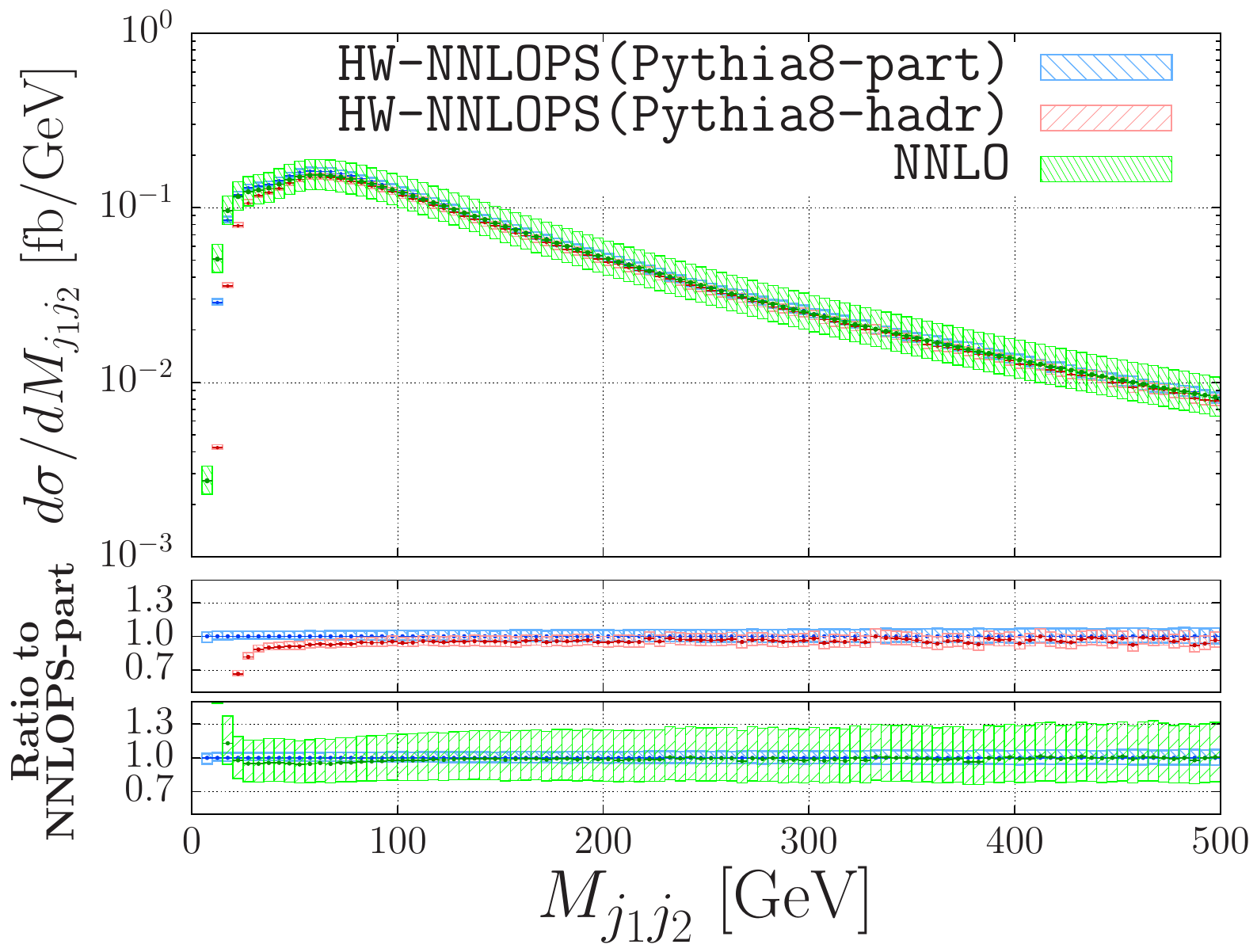}
  \caption{Rapidity difference (left) and invariant mass (right)
    of two hardest jets in \NNLO{} (green), \HWNNLOPS{} (blue) and \HWNNLOPS{} with hadronization (red).
  }
  \label{fig:dely-delphi-j1j2}
\end{figure}

We have also studied a few dijet observables. In
Fig.~\ref{fig:dely-delphi-j1j2} we present a comparison between the
various simulations for the rapidity difference (left) and the
invariant mass of the two hardest jets (right).
We can see that $\Delta y_{\sss j_1,j_2}$ displays a peak in the bin
just above $\Delta y=0.4$ which is consistent with the jet radius
($R=0.4$) we used for clustering jets. A similar peak is present also
in the distribution of the azimuthal angle between the jets $\Delta
\phi_{\sss j_1,j_2}$.
We notice that the invariant mass distribution has a peak and a
noticeable shoulder (partially washed away after hadronization) at
about 55-60 and 20-35 GeV, respectively. Their origin can be
understood from the peaks in the $\Delta y_{\sss j_1, j_2}$ and
$\Delta \phi_{\sss j_1, j_2}$ distributions. In fact the invariant
mass can be written as $M_{\sss j_1j_2} = 2 p_{\sss t,j_1}p_{\sss
  t,j_2} (\cosh \Delta y_{\sss j_1,j_2} -\cos \Delta \phi_{\sss
  j_1,j_2})$. It is easy to roughly estimate the positions of the
structures present in the $M_{\sss j_1j_2}$ plot: they correspond to
when the transverse momenta of the jets are close to the transverse
momentum cut, one of the variables ($\Delta y_{j_1,j_2}$ or $\Delta
\phi_{j_1,j_2}$) is close to its peak and the other one is integrated
over.

\begin{figure}[h]
  \centering
  {{ \includegraphics[width=0.8\textwidth,page=1]{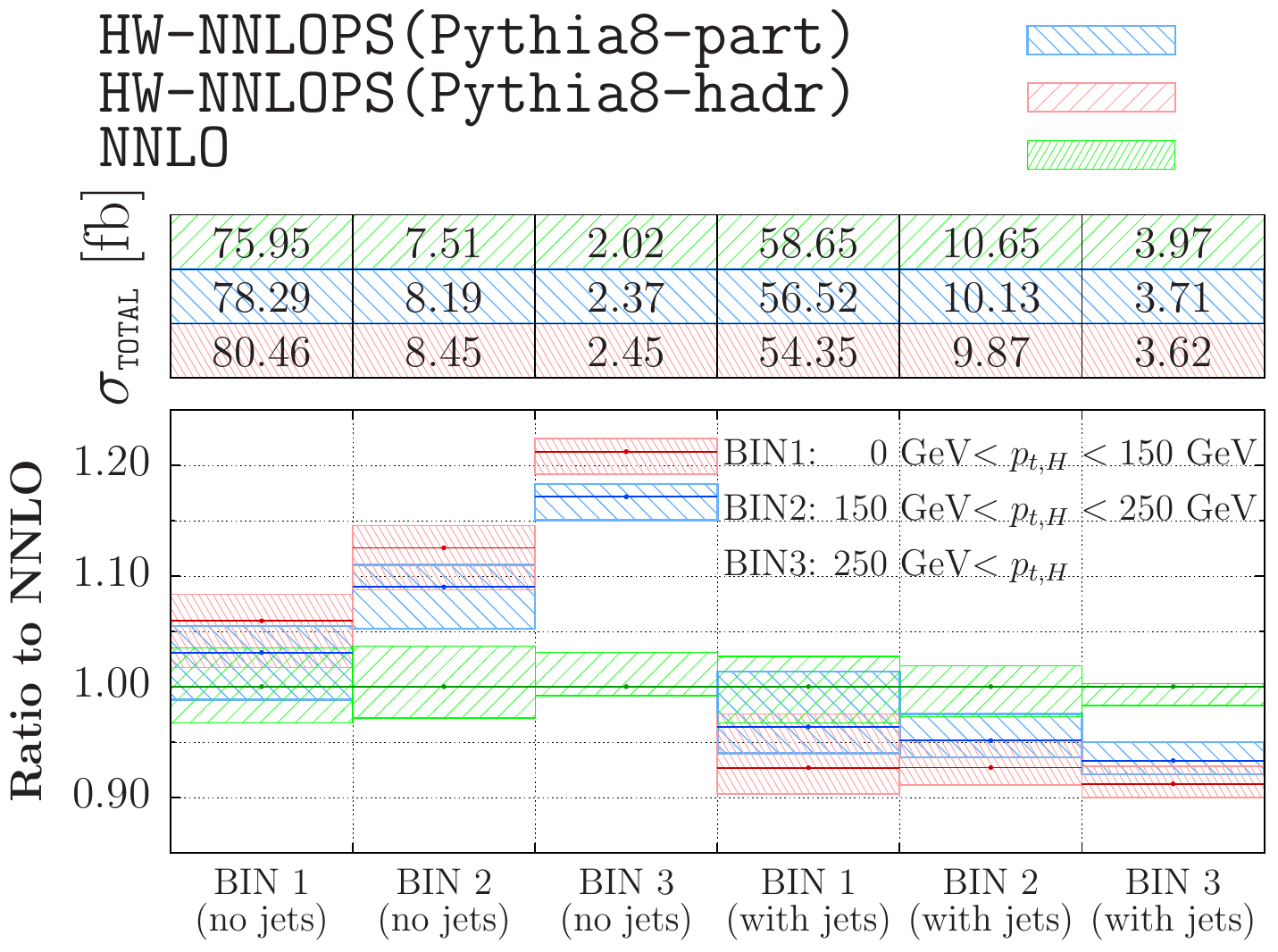}
    }}
      \caption{Total cross-section binned according to the transverse
        momentum of the Higgs boson and the presence of jets.  Jets are
        defined using the anti-$k_t$ algorithm with $R=0.4$, $p_{t,j} >
        20$ GeV and $|y_{j}| < 4.5$.  Results are shown at various levels
        of the simulation, see text for more details.  }
      \label{fig:total-xs-cuts}
\end{figure}

Finally, we examine production rates when binned into six regions
according to the transverse momentum of the Higgs boson (3 bins
corresponding to $0 < \pth<150$~GeV, $150<\pth<250$~GeV, and
$250$~GeV$ < \pth$) and the presence or absence of an additional jet
(with jet-veto or with one or more jets).
In Fig.~\ref{fig:total-xs-cuts} we show the six cross-sections, after
showering \HWNNLOLHE{} events with \PYTHIA{8} (\HWNNLOPS{}) with and without
hadronization, and the pure NNLO predictions.
We notice that, due to radiation that ends up outside the jet, jets
may be softened during parton shower evolution and hence the jet-veto
cross-sections are larger at \HWNNLOPS{} at parton level level
compared to pure NNLO level.  Differences can reach up to about 15\%
in the zero-jet bin when the Higgs boson has large transverse
momentum.
This effect is strengthened once hadronization is applied, since 
hadronization soften the leading jet spectrum even further. In this
case differences up to about 20\% can be found compared to pure NNLO
predictions.
One reason for these sizable differences between NNLO and \HWNNLOPS{}
predictions is that the jet threshold used here is relatively soft
($20$~GeV). In this region the NNLO prediction is starting to diverge
and the the leading jet transverse momentum spectrum is particularly
sensitive to soft emissions and hadronization effects, as shown in
Fig.~\ref{fig:pt-j1j2}. Furthermore, increasing the value of the jet
radius would limit the impact of out-of-jet radiation.  Nevertheless,
these numbers demonstrate that the merging NNLO calculations to parton
showers can be very important when realistic fiducial cuts are
applied.

\section{Conclusion}
\label{sec:conclu}

In this paper we have used the \MINLO{}-based merging method to
obtain the first NNLO accurate predictions for \HW{} production
consistently matched to a parton shower, including the decay of the
$\textrm{W}$ boson to leptons. The method requires a
multi-differential reweighting of the weight of \HWJMINLO{} events to
the NNLO accurate Born distributions. We have used that the
$K-$factor, within our statistical accuracy, is independent of the
mass of the dilepton system over the whole phase space, hence we have
performed the reweighting in the three Born variables $\{ \yhw{},
\Delta \yhw{}, \pth{}\}$ and in the two Collins-Soper angles that
describe the decay of the $W$ boson. For the latter variables, we have
exploited the fact that the kinematic dependence can be parametrized
in terms of spherical harmonics of degree up to two.

For our phenomenological results, we have considered a setup suggested
recently in the context of the Higgs cross section working group.  We
find that including NNLO corrections in the \MINLO{} simulation
reduces scale variation uncertainties from about 10\% to about
1-2\%. Compared to a pure NNLO calculation, while the perturbative
accuracy is the same, our tool allows one to perform fully realistic
simulations, including the study of non-perturbative effects and
multi-parton interactions.

By construction, for leptonic observables we find that the NNLO and
\NNLOPS{} simulations agree when no cut on additional radiation is
imposed.  However, we find sizable differences between the two
simulations when realistic cuts are imposed. This is particularly the
case in the region where the Higgs boson is boosted and a jet-veto
condition is imposed. In this case differences amount to about 15\% at
the 13 TeV LHC. This large effect is due to a migration of events
that, before the parton shower, have a soft jet (whose transverse
momentum is just above the veto scale) from the one-jet to the
zero-jet category. In fact, with our setup, the main effect of the
parton shower is to soften the leading jet, therefore increasing the
fraction of events that fall into the zero-jet category. Different
jet-thresholds and jet-radii leads to quite different conclusions.
Still, these differences are in general outside the scale-variation
uncertainties of the NNLO calculation, hence the \NNLOPS{} accurate
prediction becomes important to provide a more realistic uncertainty
estimate.
The \HVNNLOPS{} generator we have
developed will allow to simulate these features in a fully-exclusive
way, retaining at the same time all the virtues of an NNLO computation
for fully inclusive observables, as well as resummation effects,
thanks to the interplay among \POWHEG{}, \MINLO{} and parton
showering.

\section*{Acknowledgments}
We thank Giancarlo Ferrera and Francesco Tramontano for providing a
preliminary version of their NNLO \HVNNLO{} code and for extensive
discussion.
We are also grateful to Alexander Karlberg, Zoltan Kunszt, Paolo
Nason, and Carlo Oleari for useful exchanges.  WA, WB, and ER thank CERN for
hospitality, and WA, WB, ER and GZ would like to express a special
thanks to the Mainz Institute for Theoretical Physics (MITP) for its
hospitality and support while part of this work was carried out.
The research of WA, WB, GZ, and, in part, of ER, is supported by the
ERC grant 614577 ``HICCUP -- High Impact Cross Section Calculations
for Ultimate Precision''.
\appendix

\section{Pure NNLO Uncertainties}
\label{App:uncertainties}
This section  we compare the 49 scale method we used, as detailed in Sec.~\ref{subsec:Estimating-uncertainties}, to the 21 scale method used for \HNNLOPS{} \cite{Hamilton:2013fea} and \DYNNLOPS{} \cite{Karlberg:2014qua}. To do this we repeated our analysis using the 21 scale uncertainty method, with $(\Kr,\Kf)=(0.5,0.5),(1,1),(2,2)$ for the fixed order NNLO results.
We find that in general both methods result in uncertainty bands they are very similar, with the 49 scale uncertainty band being
only 1-2 permille larger in some bins.

\begin{figure}[h]
  \centering
      {{ \includegraphics[width=0.48\textwidth,page=1]{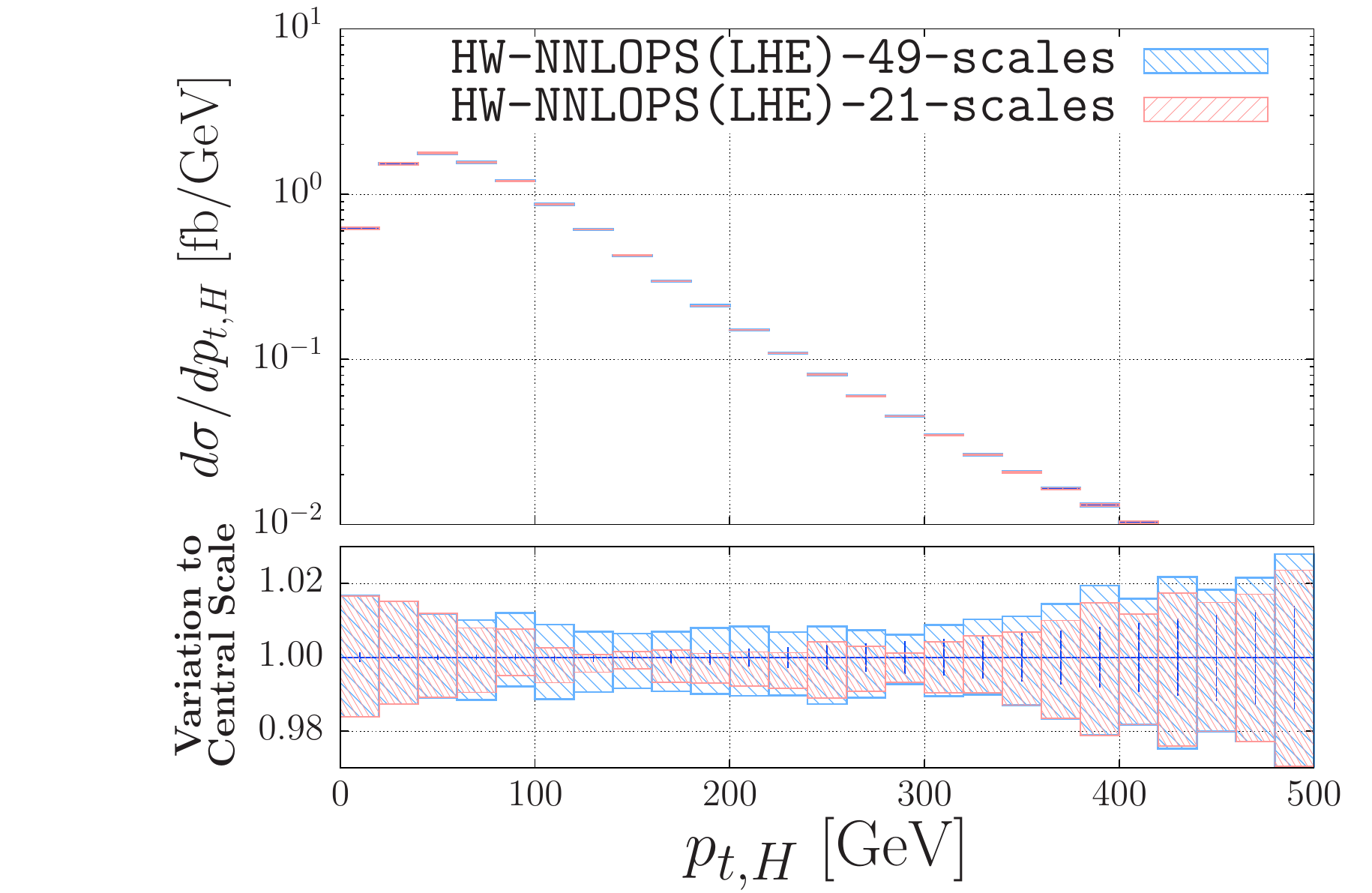} }}
      {{ \includegraphics[width=0.48\textwidth,page=1]{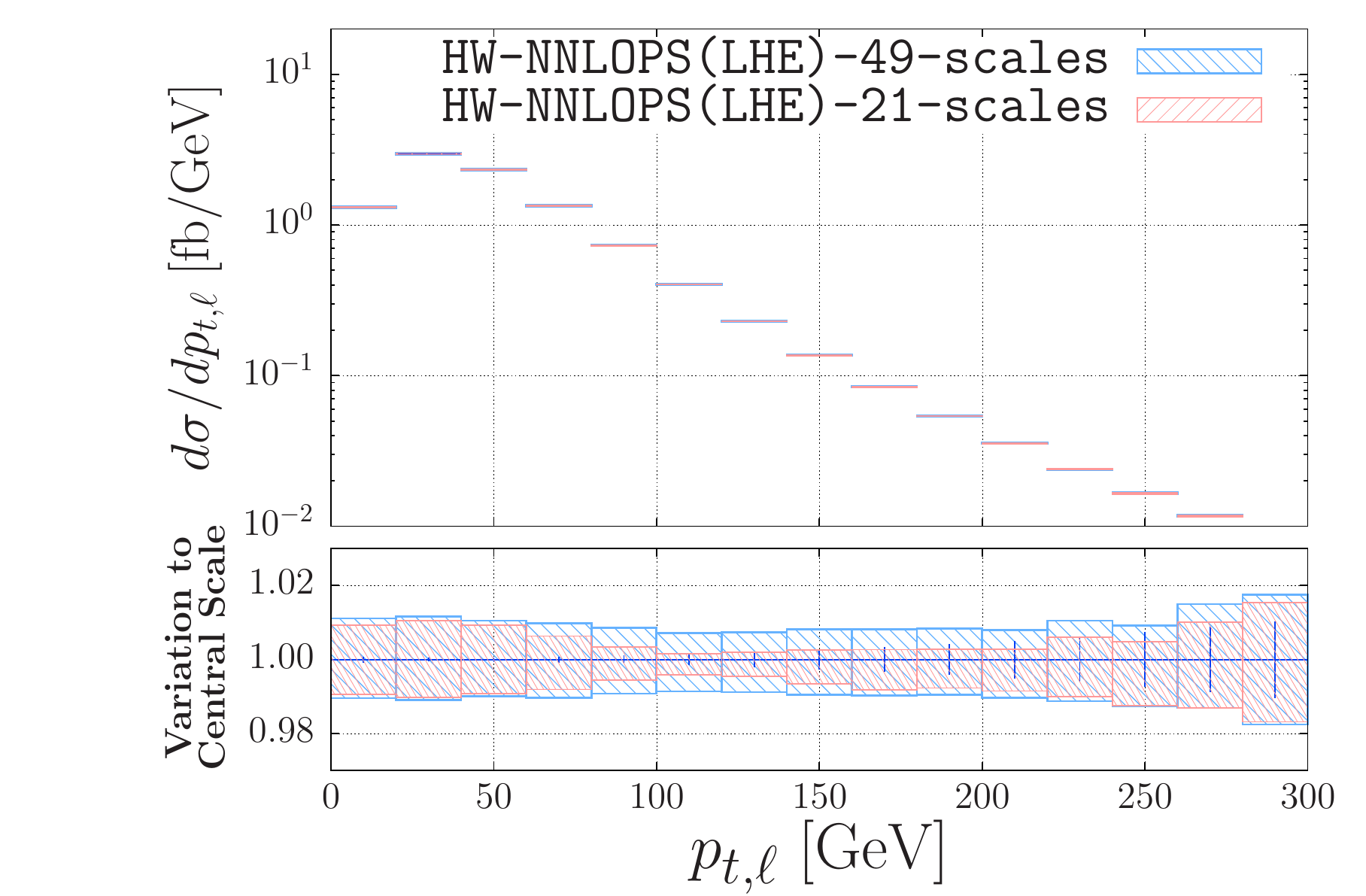} }}
      {{ \includegraphics[width=0.48\textwidth,page=1]{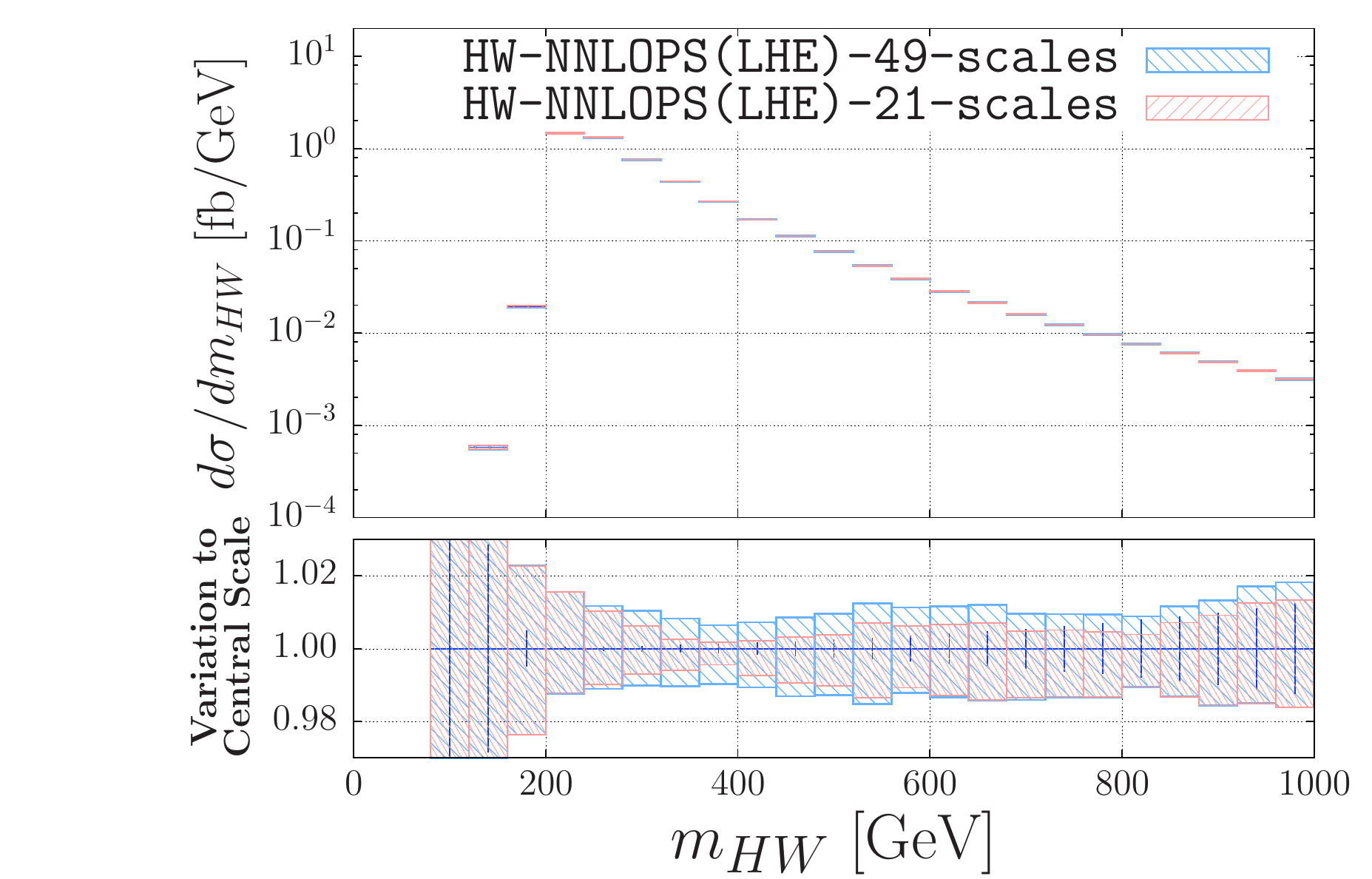} }}
      {{ \includegraphics[width=0.48\textwidth,page=1]{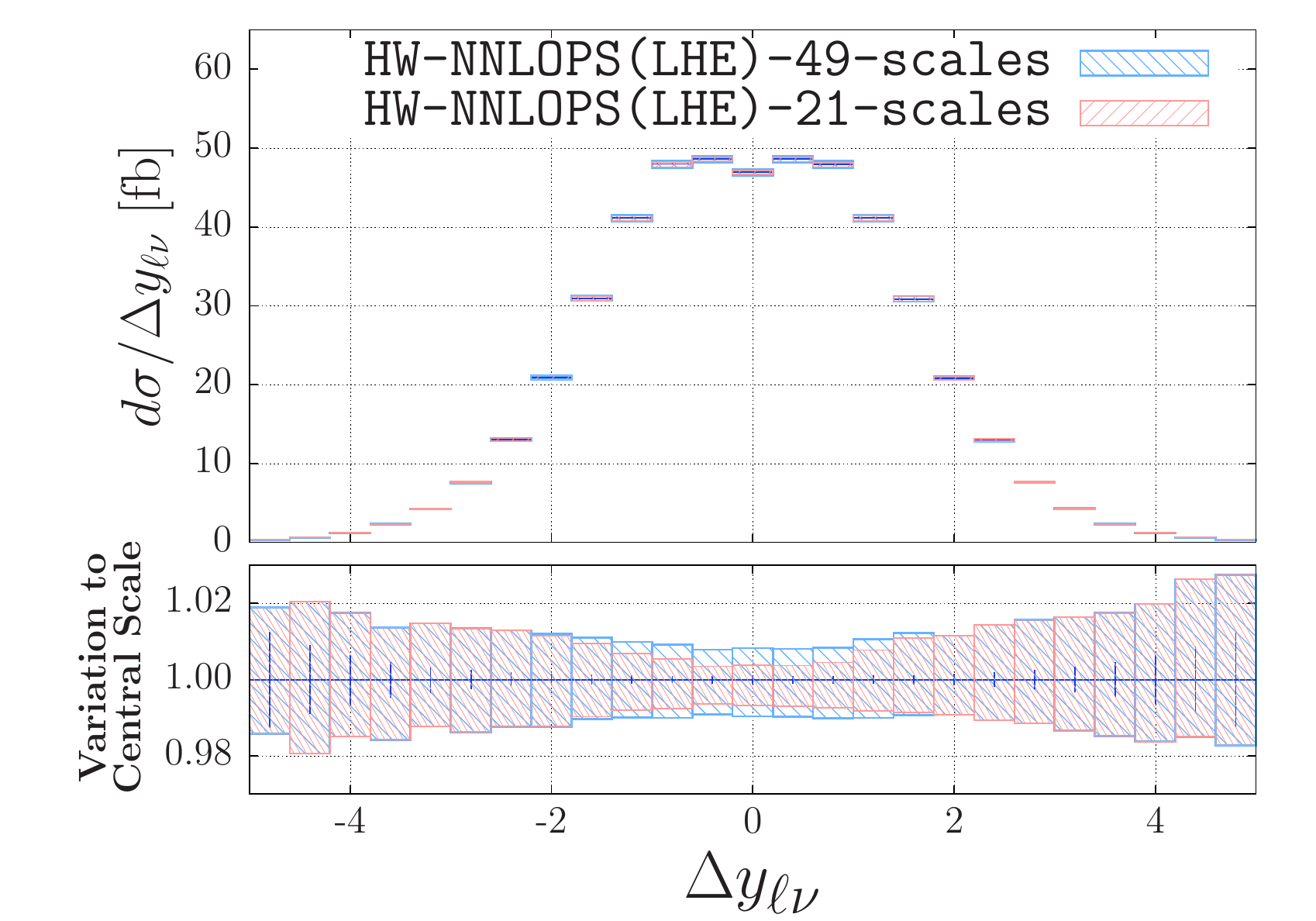} }}
      \caption{Comparison of the uncertainty from the envelope of 49 scale variations (blue) vs 21 scale variations (green) for $\pth$ (upper left), $p_{t,\sss\ell}$ (upper right), for $m_{\sss HW}$ (lower left), and for $\Delta y_{\ell\nu}$. Statistical error of the central scale result shown for reference.}
      \label{fig:scalepinch}
\end{figure}

There are however few cases where having only 21 scales results in
noticeably smaller uncertainty bands than 49 scales. 
To quantify better the differences between the two uncertainties from the two methods, we show in Fig.~\ref{fig:scalepinch} 
four observables for which we found the largest differences in uncertainties bands.

\begin{thebibliography}{99}

\bibitem{Aad:2012tfa} 
  G.~Aad {\it et al.} [ATLAS Collaboration],
  Phys.\ Lett.\ B {\bf 716}, 1 (2012)
  [\href{http://xxx.lanl.gov/abs/1207.7214}{{\tt arXiv:1207.7214}}~[hep-ex]]

\bibitem{Chatrchyan:2012xdj} 
  S.~Chatrchyan {\it et al.} [CMS Collaboration],
  Phys.\ Lett.\ B {\bf 716}, 30 (2012)
  [\href{http://xxx.lanl.gov/abs/1207.7235}{{\tt arXiv:1207.7235}}~[hep-ex]]

\bibitem{Aad:2014xzb} 
  G.~Aad {\it et al.} [ATLAS Collaboration],
  JHEP {\bf 1501}, 069 (2015)
  [\href{http://xxx.lanl.gov/abs/1409.6212}{{\tt arXiv:1409.6212}}~[hep-ex]]
  
\bibitem{Chatrchyan:2013zna} 
  S.~Chatrchyan {\it et al.} [CMS Collaboration],
  Phys.\ Rev.\ D {\bf 89}, no. 1, 012003 (2014)
  [\href{http://xxx.lanl.gov/abs/1310.3687}{{\tt arXiv:1310.3687}}~[hep-ex]]


\bibitem{TheATLAScollaboration:2013hia} 
  The ATLAS collaboration,
  ATLAS-CONF-2013-075.

\bibitem{CMS:zwa} 
  The CMS Collaboration,
  CMS-PAS-HIG-13-009.

\bibitem{CMS:ckv} 
  The CMS Collaboration,
  CMS-PAS-HIG-12-053.

\bibitem{Chatrchyan:2014tja} 
  S.~Chatrchyan {\it et al.} [CMS Collaboration],
  Eur.\ Phys.\ J.\ C {\bf 74}, 2980 (2014)
  [\href{http://xxx.lanl.gov/abs/1404.1344}{{\tt arXiv:1404.1344}}~[hep-ex]]

\bibitem{Butterworth:2008iy} 
  J.~M.~Butterworth, A.~R.~Davison, M.~Rubin and G.~P.~Salam,
  Phys.\ Rev.\ Lett.\  {\bf 100}, 242001 (2008)
  [\href{http://xxx.lanl.gov/abs/0802.2470}{{\tt arXiv:0802.2470}}~[hep-ph]]

\bibitem{Brein:2003wg} 
  O.~Brein, A.~Djouadi and R.~Harlander,
  Phys.\ Lett.\ B {\bf 579}, 149 (2004)
  [\href{http://xxx.lanl.gov/abs/hep-ph/0307206}{{\tt hep-ph/0307206}}]

\bibitem{Ferrera:2011bk} 
  G.~Ferrera, M.~Grazzini and F.~Tramontano,
  Phys.\ Rev.\ Lett.\  {\bf 107}, 152003 (2011)
  [\href{http://xxx.lanl.gov/abs/1107.1164}{{\tt arXiv:1107.1164}}~[hep-ph]]

\bibitem{Ferrera:2014lca} 
  G.~Ferrera, M.~Grazzini and F.~Tramontano,
  Phys.\ Lett.\ B {\bf 740}, 51 (2015)
  [\href{http://xxx.lanl.gov/abs/1407.4747}{{\tt arXiv:1407.4747}}~[hep-ph]]

\bibitem{Brein:2011vx} 
  O.~Brein, R.~Harlander, M.~Wiesemann and T.~Zirke,
  Eur.\ Phys.\ J.\ C {\bf 72}, 1868 (2012)
  [\href{http://xxx.lanl.gov/abs/1111.0761}{{\tt arXiv:1111.0761}}~[hep-ph]]

\bibitem{Ferrera:2013yga} 
  G.~Ferrera, M.~Grazzini and F.~Tramontano,
  JHEP {\bf 1404}, 039 (2014)
  [\href{http://xxx.lanl.gov/abs/1312.1669}{{\tt arXiv:1312.1669}}~[hep-ph]]

\bibitem{Ciccolini:2003jy} 
  M.~L.~Ciccolini, S.~Dittmaier and M.~Kramer,
  Phys.\ Rev.\ D {\bf 68}, 073003 (2003)
  [\href{http://xxx.lanl.gov/abs/hep-ph/0306234}{{\tt hep-ph/0306234}}]

\bibitem{Denner:2011id} 
  A.~Denner, S.~Dittmaier, S.~Kallweit and A.~Muck,
  JHEP {\bf 1203}, 075 (2012)
  [\href{http://xxx.lanl.gov/abs/1112.5142}{{\tt arXiv:1112.5142}}~[hep-ph]]  

\bibitem{Denner:2014cla} 
  A.~Denner, S.~Dittmaier, S.~Kallweit and A.~Muck,
  Comput.\ Phys.\ Commun.\  {\bf 195}, 161 (2015)
  [\href{http://xxx.lanl.gov/abs/1412.5390}{{\tt arXiv:1412.5390}}~[hep-ph]]  

  
\bibitem{Campbell:2016jau}
  J.~M.~Campbell, R.~K.~Ellis and C.~Williams,
  [\href{http://xxx.lanl.gov/abs/1601.00658}{{\tt arXiv:1601.00658}}~[hep-ph]]

\bibitem{Banfi:2012jm} 
  A.~Banfi, P.~F.~Monni, G.~P.~Salam and G.~Zanderighi,
  Phys.\ Rev.\ Lett.\  {\bf 109}, 202001 (2012)
  [\href{http://xxx.lanl.gov/abs/1206.4998}{{\tt arXiv:1206.4998}}~[hep-ph]]

\bibitem{Becher:2014aya} 
  T.~Becher, R.~Frederix, M.~Neubert and L.~Rothen,
  Eur.\ Phys.\ J.\ C {\bf 75}, no. 4, 154 (2015)
  [\href{http://xxx.lanl.gov/abs/1412.8408}{{\tt arXiv:1412.8408}}~[hep-ph]]

\bibitem{Frixione:2002ik} 
  S.~Frixione and B.~R.~Webber,
  JHEP {\bf 0206}, 029 (2002)
  [\href{http://xxx.lanl.gov/abs/hep-ph/0204244}{{\tt hep-ph/0204244}}]

\bibitem{Nason:2004rx} 
  P.~Nason,
  JHEP {\bf 0411}, 040 (2004)
  [\href{http://xxx.lanl.gov/abs/hep-ph/0409146}{{\tt hep-ph/0409146}}]

\bibitem{Frixione:2005gz} 
  S.~Frixione and B.~R.~Webber
  [\href{http://xxx.lanl.gov/abs/hep-ph/0506182.}{{\tt hep-ph/0506182}}]


\bibitem{Luisoni:2013kna} 
  G.~Luisoni, P.~Nason, C.~Oleari and F.~Tramontano,
  JHEP {\bf 1310}, 083 (2013)
  [\href{http://xxx.lanl.gov/abs/1306.2542}{{\tt arXiv:1306.2542}}~[hep-ph]]

\bibitem{Goncalves:2015mfa}
  D.~Goncalves, F.~Krauss, S.~Kuttimalai and P.~Maierh\"ofer,
  Phys.\ Rev.\ D {\bf 92} (2015) 7,  073006
  [\href{http://xxx.lanl.gov/abs/1509.01597}{{\tt arXiv:1509.01597}}~[hep-ph]]


\bibitem{Hamilton:2012np} 
  K.~Hamilton, P.~Nason and G.~Zanderighi,
  JHEP {\bf 1210}, 155 (2012)
  [\href{http://xxx.lanl.gov/abs/1206.3572}{{\tt arXiv:1206.3572}}~[hep-ph]]

\bibitem{Hamilton:2012rf} 
  K.~Hamilton, P.~Nason, C.~Oleari and G.~Zanderighi,
  JHEP {\bf 1305}, 082 (2013)
  [\href{http://xxx.lanl.gov/abs/1212.4504}{{\tt arXiv:1212.4504}}~[hep-ph]]

\bibitem{Hamilton:2013fea} 
  K.~Hamilton, P.~Nason, E.~Re and G.~Zanderighi,
  JHEP {\bf 1310}, 222 (2013)
  [\href{http://xxx.lanl.gov/abs/1309.0017}{{\tt arXiv:1309.0017}}~[hep-ph]]

\bibitem{Karlberg:2014qua} 
  A.~Karlberg, E.~Re and G.~Zanderighi,
  JHEP {\bf 1409}, 134 (2014)
  [\href{http://xxx.lanl.gov/abs/1407.2940}{{\tt arXiv:1407.2940}}~[hep-ph]]

\bibitem{Frederix:2015fyz}
  R.~Frederix and K.~Hamilton,
  [\href{http://xxx.lanl.gov/abs/1512.02663}{{\tt arXiv:1512.02663}}~[hep-ph]]


\bibitem{Collins:1977iv}
  J.~C.~Collins and D.~E.~Soper,
  Phys.\ Rev.\ D {\bf 16} (1977) 2219.

\bibitem{Catani:1993hr}
  S.~Catani, Y.~L.~Dokshitzer, M.~H.~Seymour and B.~R.~Webber,
  Nucl.\ Phys.\ B {\bf 406} (1993) 187.
  
\bibitem{Ellis:1993tq}
  S.~D.~Ellis and D.~E.~Soper,
  Phys.\ Rev.\ D {\bf 48} (1993) 3160
  [\href{http://xxx.lanl.gov/abs/hep-ph/9305266}{{\tt hep-ph/9305266}}]
  
\bibitem{hvnnlo}
  G. Ferrera,  M. Grazzini, F. Tramontano, private communication. 
  
  

   
\bibitem{Alioli:2010xd}
  S.~Alioli, P.~Nason, C.~Oleari and E.~Re,
  JHEP {\bf 1006} (2010) 043
  [\href{http://xxx.lanl.gov/abs/1002.2581}{{\tt arXiv:1002.2581}}~[hep-ph]]
  
\bibitem{Harland-Lang:2014zoa}
  L.~A.~Harland-Lang, A.~D.~Martin, P.~Motylinski and R.~S.~Thorne,
  Eur.\ Phys.\ J.\ C {\bf 75} (2015) 5,  204
  [\href{http://xxx.lanl.gov/abs/1412.3989}{{\tt arXiv:1412.3989}}~[hep-ph]]


\bibitem{Cacciari:2008gp}
  M.~Cacciari, G.~P.~Salam and G.~Soyez,
  JHEP {\bf 0804} (2008) 063
  [\href{http://xxx.lanl.gov/abs/0802.1189}{{\tt arXiv:0802.1189}}~[hep-ph]]


\bibitem{Cacciari:2005hq}
  M.~Cacciari and G.~P.~Salam,
  Phys.\ Lett.\ B {\bf 641} (2006) 57
  [\href{http://xxx.lanl.gov/abs/hep-ph/0512210}{{\tt hep-ph/0512210}}]
  
\bibitem{Cacciari:2011ma}
  M.~Cacciari, G.~P.~Salam and G.~Soyez,
  Eur.\ Phys.\ J.\ C {\bf 72} (2012) 1896
  [\href{http://xxx.lanl.gov/abs/1111.6097}{{\tt arXiv:1111.6097}}~[hep-ph]]

\bibitem{Sjostrand:2007gs}
  T.~Sjostrand, S.~Mrenna and P.~Z.~Skands,
  Comput.\ Phys.\ Commun.\  {\bf 178} (2008) 852
  [\href{http://xxx.lanl.gov/abs/0710.3820}{{\tt arXiv:0710.3820}}~[hep-ph]]


\bibitem{Skands:2014pea}
  P.~Skands, S.~Carrazza and J.~Rojo,
  Eur.\ Phys.\ J.\ C {\bf 74} (2014) 8,  3024
  [\href{http://xxx.lanl.gov/abs/1404.5630}{{\tt arXiv:1404.5630}}~[hep-ph]]
  
\bibitem{Nason:2013uba}
  P.~Nason and C.~Oleari,
  [\href{http://xxx.lanl.gov/abs/1303.3922}{{\tt arXiv:1303.3922}}~[hep-ph]]
  
\bibitem{Banfi:2012yh}
  A.~Banfi, G.~P.~Salam and G.~Zanderighi,
  JHEP {\bf 1206} (2012) 159
  [\href{http://xxx.lanl.gov/abs/1203.5773}{{\tt arXiv:1203.5773}}~[hep-ph]]
  

\bibitem{Alioli:2008tz}
  S.~Alioli, P.~Nason, C.~Oleari and E.~Re,
  JHEP {\bf 0904} (2009) 002
  [\href{http://xxx.lanl.gov/abs/0812.0578}{{\tt arXiv:0812.0578}}~[hep-ph]]



\end{thebibliography}
\end{document}